%% file: DocumentForgery.tex
\documentclass[journal]{IEEEtran}

\ifCLASSINFOpdf
\else
\fi

\usepackage[cmex10]{amsmath}
\usepackage{amssymb}
\usepackage{gensymb}
\usepackage{textcomp}
\usepackage{pifont}
\usepackage{graphics}
\usepackage{cite}
\usepackage[OT2,T1]{fontenc}
\usepackage{subfigure}
\usepackage{url}

\usepackage{float}
\usepackage{amsmath}
\usepackage{amstext}
\usepackage{amsfonts}
\usepackage[pdftex]{graphicx}
\usepackage{makecell}
\usepackage{array,multirow}
\usepackage{multicol}
\usepackage{color}

\usepackage{mathtools}
\usepackage{booktabs,tabularx}

\newcolumntype{C}[1]{>{\centering\arraybackslash}p{#1}}

\usepackage{flushend}

\begin{document}
%
\title{Deep Learning-based Forgery Attack on\\Document Images}
%
%
%

\author{Lin Zhao,~\IEEEmembership{Student~Member,~IEEE},
        Changsheng Chen,~\IEEEmembership{Senior Member,~IEEE},
        and~Jiwu Huang,~\IEEEmembership{Fellow,~IEEE}

\thanks{The authors are with the Guangdong Key Laboratory of Intelligent Information Processing and Shenzhen Key Laboratory of Media Security, and National Engineering Laboratory for Big Data System Computing Technology, College of Electronics and Information Engineering, Shenzhen University, Shenzhen, China. They are also with Shenzhen Institute of Artificial Intelligence and Robotics for Society, China (e-mail: zhaolin2016@email.szu.edu.cn, cschen@szu.edu.cn, jwhuang@szu.edu.cn).}
}

\maketitle

\begin{abstract}
With the ongoing popularization of online services, the digital document images have been used in various applications.
Meanwhile, there have emerged some deep learning-based text editing algorithms which alter the textual information of an image in an end-to-end fashion.
In this work, we present a low-cost document forgery algorithm by the existing deep learning-based technologies to edit practical document images.
To achieve this goal, the limitations of existing text editing algorithms towards complicated characters and complex background are addressed by a set of network design strategies. 
First, the unnecessary confusion in the supervision data is avoided by disentangling the textual and background information in the source images.
Second, to capture the structure of some complicated components, the text skeleton is provided as auxiliary information and the continuity in texture is considered explicitly in the loss function.
Third, the forgery traces induced by the text editing operation are mitigated by some post-processing operations which consider the distortions from the print-and-scan channel.
Quantitative comparisons of the proposed method and the exiting approach have shown the advantages of our design by reducing the about 2/3 reconstruction error measured in MSE, improving reconstruction quality measured in PSNR and in SSIM by 4 dB and 0.21, respectively.
Qualitative experiments have confirmed that the reconstruction results of the proposed method are visually better than the existing approach in both complicated characters and complex texture.
More importantly, we have demonstrated the performance of the proposed document forgery algorithm under a practical scenario where an attacker is able to alter the textual information in an identity document using only one sample in the target domain.
The forged-and-recaptured samples created by the proposed text editing attack and recapturing operation have successfully fooled some existing document authentication systems.
\end{abstract}


\IEEEpeerreviewmaketitle
\section{Introduction}
\label{sec:Introduction}

\begin{figure}[t]
\centering
\begin{minipage}{0.45\linewidth}
\subfigure[]{
\includegraphics[width=1\textwidth]{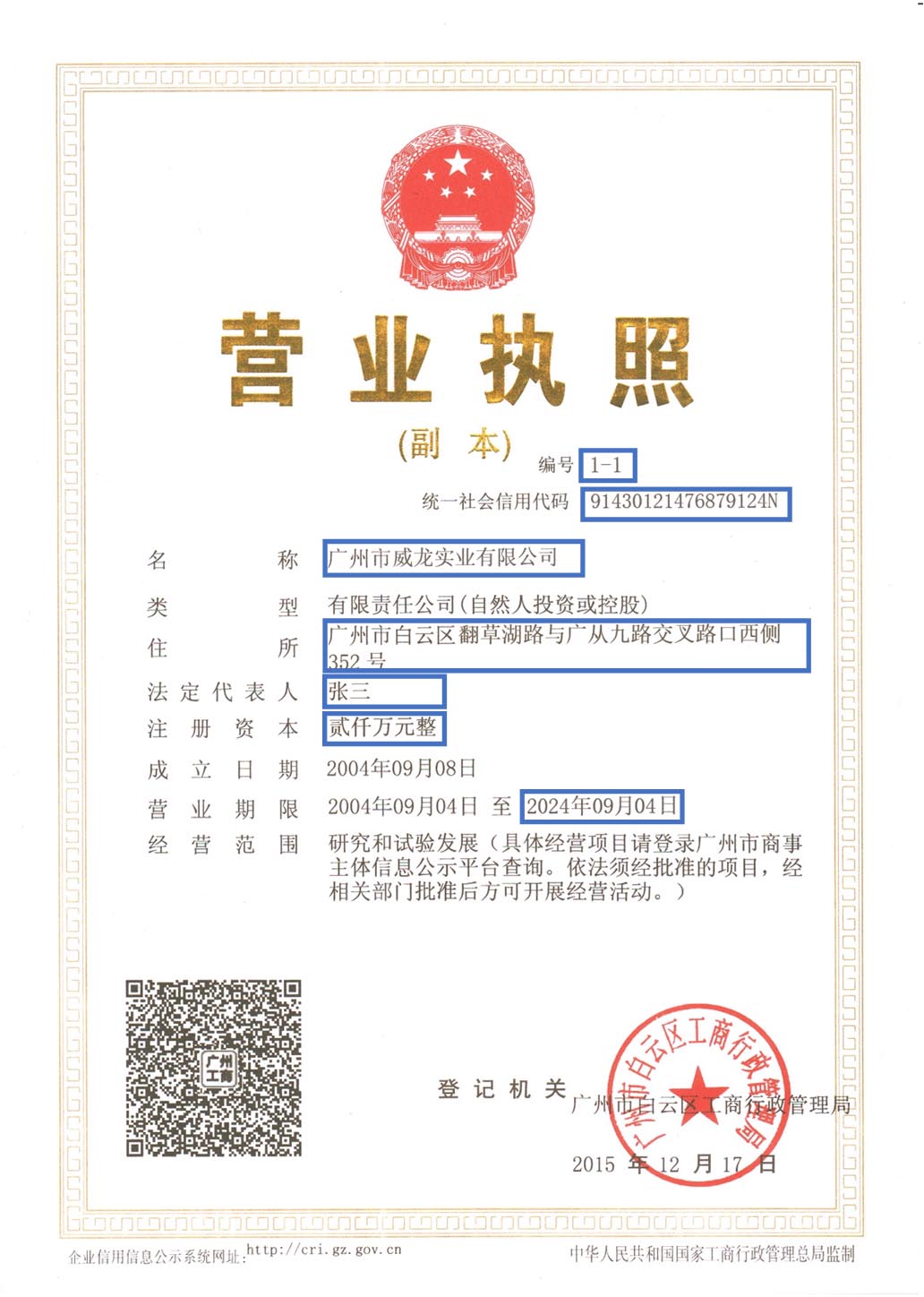}
\label{subfig:license}
}
\end{minipage}
\begin{minipage}{0.45\linewidth}\vspace{1.3mm}
\subfigure[]{
\includegraphics[width=0.9\textwidth]{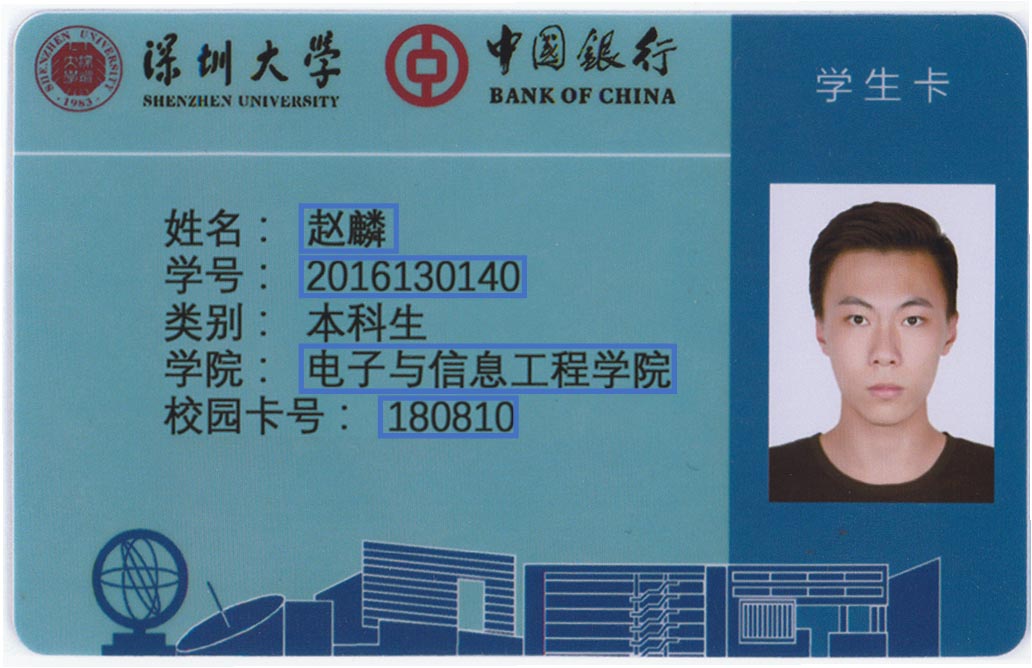}
\label{subfig:student_ID}
}\vspace{1mm}
\\
\subfigure[]{
\includegraphics[width=0.9\textwidth]{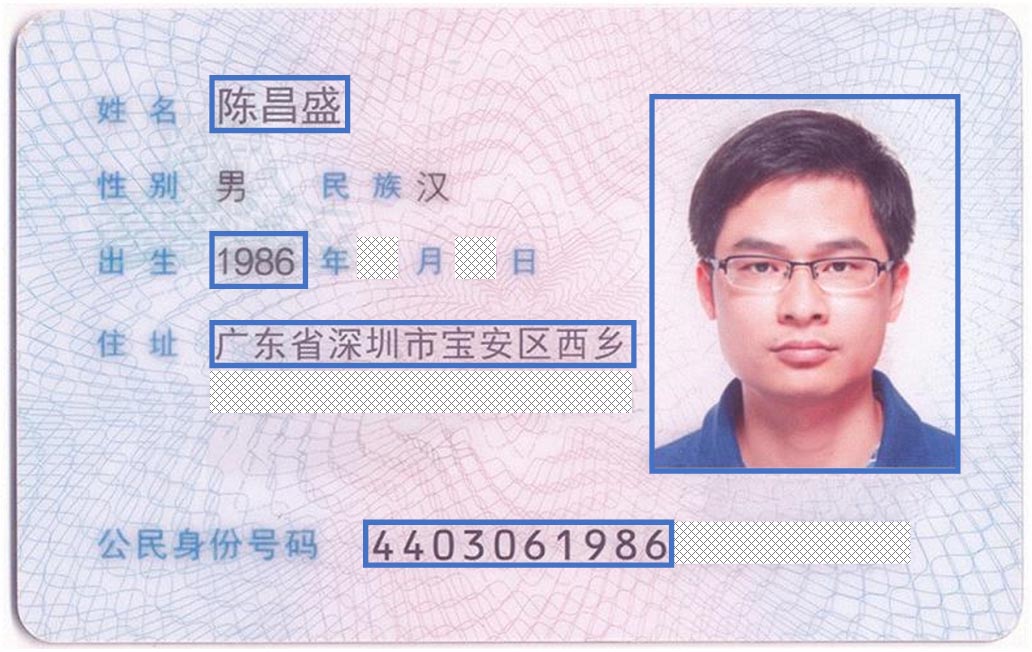}
\label{subfig:ID_card}
}
\end{minipage}
\caption{Illustration of three types of document images processed by the proposed document forgery approach (ForgeNet, as outlined in Sec.~\ref{sec:ForgeNet}). The edited regions are boxed out in blue.}
\label{fig:ID_examples}
\end{figure}

Due to the COVID-19 pandemic, we have observed an unprecedented demand for online document authentication in the applications of e-commerce and e-government.
Some important document images were uploaded to online platforms for various purposes.
However, the content of document can be altered by some image editing tools or deep learning-based technologies.
As an illustration in Fig.~\ref{subfig:license}, we show an example on Document Forgery Attack dataset from Alibaba Tianchi Competition \cite{AliTianchi} forged with the proposed document forgery approach.
Some key information on the original image is edited and then the document is recaptured to conceal the forgery trace.
It is a low-cost (automatic, and without the need of skilled professional) and dangerous act if an attacker uses such forge-and-recapture document images to launch illegal attack.


Recently, it has been demonstrated that characters and words in natural images can be edited with convolutional neural networks \cite{wu2019editing, roy2020stefann, yang2020swaptext} in an end-to-end fashion.
Similar to the framework of DeepFake\cite{korshunov2018deepfakes}, these models have been trained to disentangle different components in the document images, such as text, style and background.
During the process of text editing, the input textual information (plain text with the targeted contents) is converted to a text image with targeted style and background.
It should be noted that these works \cite{wu2019editing, roy2020stefann, yang2020swaptext} are originally proposed for the visual translation and AR translation applications.
To the best of our knowledge, there is no existing works on evaluating impacts of the above deep learning-based textual contents generation schemes towards document security.
The edited text images have not been investigated from a forensic aspect.

\begin{figure*}[t!]
\begin{center}
\includegraphics[width=0.9\textwidth]{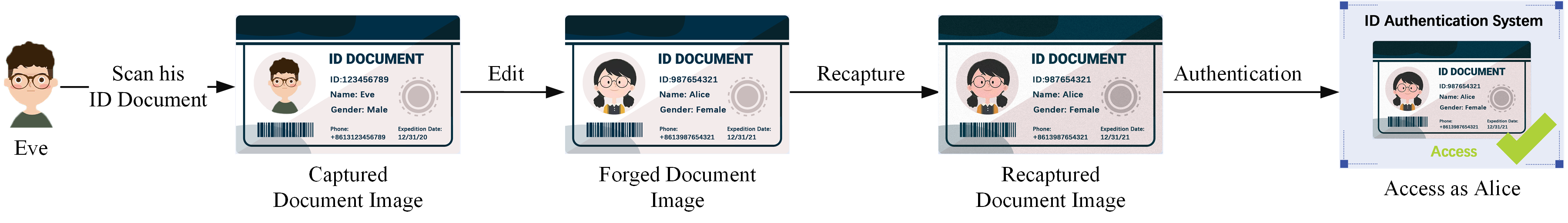}
\centerline{\footnotesize(a)}
\includegraphics[width=0.9\textwidth]{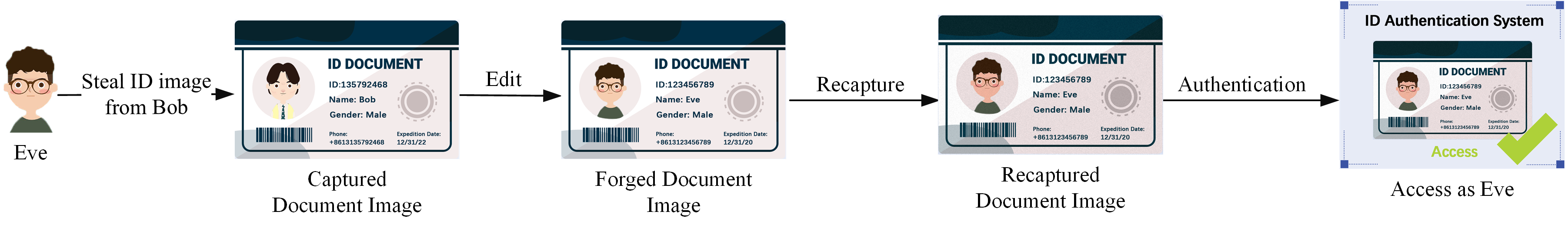}
\centerline{\footnotesize(b)}
\end{center}
\vspace{-0.25cm}
\caption{Two representative forge-and-recapture attack scenarios. (a) The attacker scans his/her own identity document to obtain an identity document image and forges the document of a target identity to perform an impersonate attack. (b) The attacker steals an identity document image and forge his/her own document to obtain unauthorized access.}
\label{fig:TreadModel}
\end{figure*}

Authentication of hardcopy documents with digitally acquired document images is a forensic research topic with broad interest.
Although the edited document image in digital domain can be traced with some existing tamper detection and localization schemes \cite{shang2015document}, it has been shown that detection of document forgery with small manipulation region (e.g., key information in a document) is challenging \cite{abramova2016detecting}.
Moreover, recapturing operation (replay attack) is an effective way to conceal the forgery trace \cite{thongkamwitoon2015image, agarwal2018diverse}.
A formal attack model with two scenarios is shown in Fig.~\ref{fig:TreadModel}.
For a common document (e.g., an identity card), the attacker's own copy can be edited to perform an impersonate attack of a target identity.
For a document with specific template, the attacker would steal a digital copy of the document, and forge his/her own document image to get unauthorized access.

To understand the security threat, one should note that detecting recapturing attack in digital documents is very different from detecting spoofing in other media, e.g., face and natural images.
For example, the forensic trace from depth in face \cite{li2018learning, chen2019attention} and natural images \cite{li2017image, agarwal2018diverse}, as well as Moir\'{e} pattern \cite{garcia2015face} artifacts in displayed images are not available in document images.
Both the captured and recaptured versions of a hardcopy document are acquired from flat paper surfaces, which lack the distinct differences between a 3D natural scene versus a flat surface or a pixelated display.
Thus, the advancement of the deep learning technologies in text editing may have already put our document image at stake.

In this work, we build a deep learning-based document forgery network to attack the existing digital document authentication system under a practical scenario.
The approach can be divided into two stages, i.e., document forgery and document recapturing.
In the document forgery stage, the target text region is disentangled to yield the text, style and background components.
To allow text editing of characters with complicated structure under complex background, several important strategies are introduced.
First, to avoid confusions in different components of the source images (e.g., between complex background textures and texts), the textual information is extracted by subsequently performing inpainting and differentiation on the input image.
Second, to capture the structure of some complicated components, the text skeleton is provided as auxiliary information and the continuity in texture is considered explicitly in the loss function.
Last but not least, the forgery traces between the forgery and background regions are mitigated by post-processing operations with considerations on distortions from the print-and-scan process.
In the recapturing stage, the forged document is printed and scanned with some off-the-shelf devices.
In the experiment, the network is trained with a publicly available document image dataset and some synthetic textual images with complicated background.
Ablation study shows the importance of our strategies in designing and training our document forgery network.
Moreover, we demonstrate the document forgery performance under a practical scenario where an attacker generates a forgery document with only one sample in the target domain.
In our case, an identity document with complex background can also be edited by a single sample fine-tuning operation.
Finally, the edited images are printed and scanned to conceal the forgery traces.
We show that the forge-and-recapture samples by the proposed attack have successfully fooled some existing document authentication systems.

The main contributions of this work are summarized as follows.
\begin{itemize}
\item We propose the first deep learning-based text editing network towards document images with complicated characters and complex background. Together with the recapturing attack, we show that the forge-and-recapture samples have successfully fooled some state-of-the-art document authentication systems.
\item We mitigate the visual artifacts introduced by the text editing operation by color pre-compensation and inverse halftoning operations, which consider the distortions from print-and-scan channel, to produce a high-quality forgery result.
\item We demonstrate the document forgery performance under a practical scenario where an attacker alters the textual information in an identity document (with Chinese characters and complex texture) by fine-tuning the proposed scheme fine-tuned with one sample in the target domain.
\end{itemize}

The remaining of this paper is organized as follows. Section II reviews the related literatures on deep learning-based text editing. Section III introduces the proposed document forgery method. Section IV describes the datasets and training procedure of our experiments. Section V compares the proposed algorithm with the exiting text editing methods, and demonstrates the feasibility of attacking the existing document authentication systems with the forge-and-recapture attack. Section VI concludes this paper.

\section{Related Work}
\label{sec:Literature Review}

Recently, text image synthesis has become a hot topic in the field of computer vision.
Text synthesis tasks have been implemented on scene images for visual translation and augmented reality applications.
The GAN-based text synthesis technique renders more realistic text regions in natural scene images.
Wu \textit{et al.} first addressed the problem of word or text-line level scene text editing by an end-to-end trainable Style Retention Network (SRNet) \cite{wu2019editing}.
SRNet consists of three learnable modules, including text conversion module, background inpainting module and fusion module, which are used for text editing, background erasure, as well as text and background fusion, respectively.
The design of the network facilitates the modules to be pre-trained separately, reduces the difficulty in end-to-end training of complicate network.
Compared with the work of character replacement, SRNet works in word-level which is a more efficient and intuitive way of document editing.
Experimental results show that SRNet is able to edit the textual information in some natural scene images.
Roy \textit{et al.} \cite{roy2020stefann} designed a Scene Text Editor using Font Adaptive Neural Network (STEFANN) to edit texts in scene images.
However, a one-hot encoding of length 26 of the target character is adopted in STEFANN to represent the 26 upper-case English alphabets in the latent feature space.
Such one-hot encoding is expandable to lower-case English alphabets and Arabic numerals.
However, it is not applicable to Chinese which is with a much larger character set (more than 3000 characters in common use) \cite{zhang2017encoding}.
Thus, STEFANN is not suitable for editing Chinese documents.
Yang \textit{et al.} \cite{yang2020swaptext} proposed an image texts swapping scheme (SwapText) in scenes with special attention on the performance in perspective and curved text images.
In the following, we mainly focus on SRNet \cite{wu2019editing} since it is the most relevant work to our task on editing text in document images for two reasons.
First, it is applicable to Chinese character unlike STEFFANN \cite{roy2020stefann}.
Second, it keeps a relatively simple network structure compared to SwapText \cite{yang2020swaptext} which considers curved texts that uncommonly found on a document.

\begin{figure}[t!]
\centering
\subfigure[]{
\centering
\includegraphics[width=0.28\linewidth]{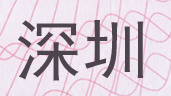}
}
\subfigure[]{
\centering
\includegraphics[width=0.28\linewidth]{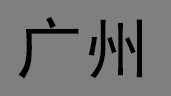}
}
\subfigure[]{
\centering
\includegraphics[width=0.28\linewidth]{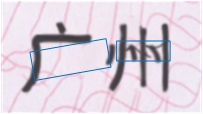}
}
\caption{Text editing under complex background by SRNet \cite{wu2019editing} (fine-tuned by 5,000 and 20,000 Chinese character images with complex and simple background, respectively). (a) Source image with styled text and background. (b) Target characters. (c) Target styled characters with background. Text artifacts and background discontinuities can be found in the boxed region of (c).}
\label{fig:BG_Limitation}
\end{figure}

The difficulties of editing Chinese text in documents images mainly lies in background inpainting and text style conversion.
In the background inpainting process, we need to fill the background after erasing the textual region.
The image background, as an important visual cue, is the main factor affecting the similarity between the synthesized and the ground-truth text images.
However, as shown in Fig.~\ref{fig:BG_Limitation}, the reconstructed regions show discontinuity in texture that degrades the visual quality.
This is mainly due to the background reconstruction loss of SRNet compares the inpainted and original images pixel by pixel and weights the distortions in different region equally, while human inspects the results mainly from the structural components, e.g., texture.

\begin{figure}[t!]
\centering
\subfigure[]{
\centering
\includegraphics[width=0.28\linewidth]{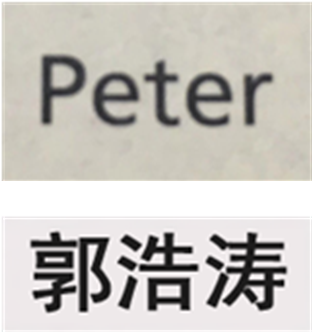}
}
\subfigure[]{
\centering
\includegraphics[width=0.28\linewidth]{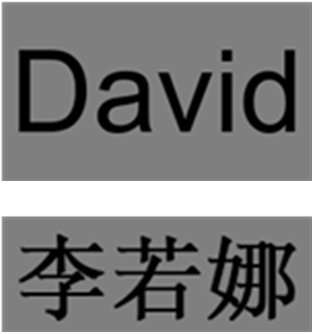}
}
\subfigure[]{
\centering
\includegraphics[width=0.28\linewidth]{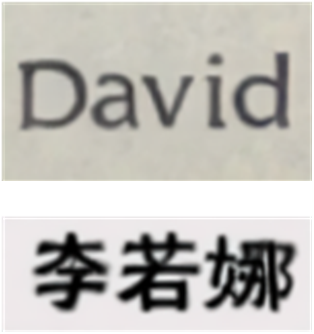}
}
\caption{Text editing of complicated Chinese characters by SRNet \cite{wu2019editing} (fine-tuned by 20,000 Chinese character images with solid color background). (a) Source image with styled text and background. (b) Target characters. (c) Target styled characters with background. The edited text languages are English (top row) and Chinese (bottom row). It should be noted that Chinese text image synthesis performs worse than English.}
\label{fig:Chinese_Limitation}
\end{figure}

In text style conversion process, the SRNet inputs the source image (with source text, target style and background) to the text conversion subnet.
However, as shown in Fig.~\ref{fig:Chinese_Limitation}(c), the text style has not been transferred from (a) to (c).
Especially, the Chinese character with more strokes is distorted more seriously than the English alphabets.
This is because different components (source text, target style, and background) in the source image introduces confusion in the text style conversion process.
It should be noted that such distortion is more obvious for Chinese characters due to two reasons.
On the one hand, the number of Chinese characters is huge, with more than 3,000 characters in common use.
It is more difficult to train a style conversion network for thousands of Chinese characters than dozens of English alphabets.
On the other hand, the font composition of Chinese characters is complex, as it consists of five standard strokes with multiple radicals.
Therefore, text editing of Chinese characters in document with complex background still presents great challenges.

In addition, most of the target contents of the existing works are scene images rather than document images.
It requires the artifacts in synthesized text image to be unobtrusive towards human visual system, rather than undetectable under forensic tools.
Therefore, the existing works \cite{wu2019editing, roy2020stefann, yang2020swaptext} have not considered to further process the text editing results with regards to the distortions from print-and-scan channel, such as color degradation, and halftoning \cite{zhang2018accurate}.

\begin{figure}[t!]
\centerline{\includegraphics[width=\linewidth]{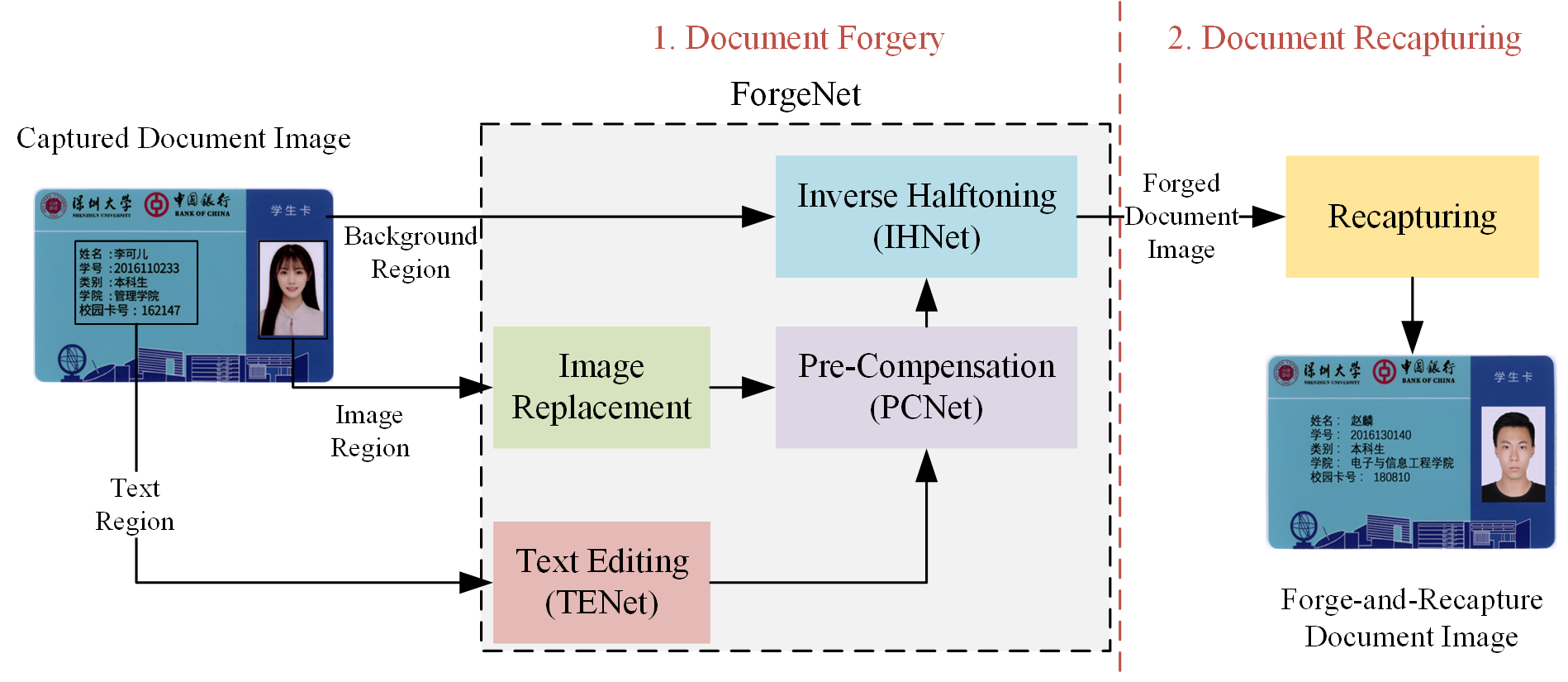}}
\caption{Overview of the proposed document forgery approach. A forge-and-recapture attack is performed on a captured document image.}
\label{fig:ForgeNet}
\end{figure}

\begin{figure*}[t!]
\centerline{\includegraphics[width=0.95\textwidth]{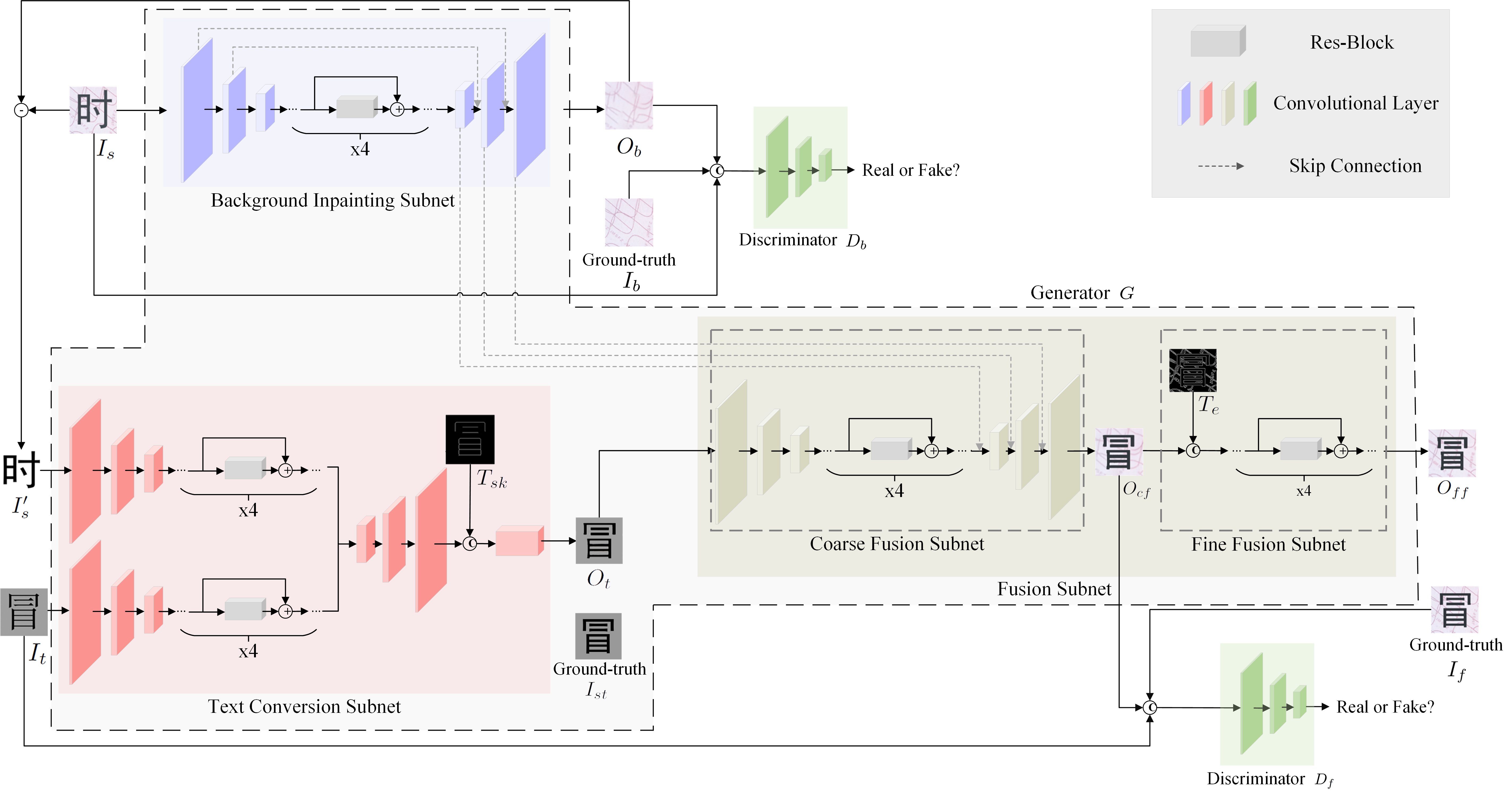}}
\caption{The framework of TENet. It contains three subnets: background inpainting subnet, text conversion subnet and fusion subnet. The background inpainting subnet generates a complete background by filling the original text region with the predicted content. The text conversion subnet replaces the text content of the source image with the target text while preserving the original style. The fusion subnet merges the output from the last two subnets and yields the edited image with the target text and original background.}
\label{fig:TENet}
\end{figure*}

\section{Proposed Method}
\label{sec:ForgeNet}

As shown in Fig.~\ref{fig:ForgeNet}, the document forgery attack is divided into the forgery (through the proposed deep network, ForgeNet) and recapturing steps.
For the forgery process, the document image acquired by an imaging device is employed as input to the ForgeNet.
It is divided into three regions, i.e., text region, image region, and background region (the areas that are not included in the first two categories).
The background region is processed by the inverse halftoning module (IHNet) to remove the halftone dots in the printed document.
The original content in the image region is replaced by the target image, and the resulting image is fed into the print-and-scan pre-compensation module (PCNet) and IHNet.
It should be noted that the PCNet deliberately distorts the color and introduces halftone patterns in the edited region such that the discrepancies between the edited and background regions are compensated.
The text region is subsequently forwarded to the text editing module (TENet), PCNet and IHNet.
After processed by the ForgeNet, the three regions are stitched together to form a complete document image.
Lastly, the forged document image is recaptured by cameras or scanners to finish the forge-and-recapture attack.
For clarity, the definitions of main symbols in our work is summarized in Tab.~\ref{tab:SymbolDefinition}.
In the following paragraphs, the TENet, PCNet, and IHNet within the ForgeNet will be elaborated.

\subsection{The Text Editing Network (TENet)}
\label{subsec:TextEdit}

In this part, a deep learning-based architecture, TENet is proposed to edit the textual information in document images.
As shown in Fig.~\ref{fig:TENet}, TENet consists of three subnets.
The background inpainting subnet generates a complete background by filling the original text region with the predicted content.
The text conversion subnet replaces the text content of the source image $I_s$ with the target text $I_t$ while preserving the original style.
The fusion subnet merges the output from the last two subnets and yields the edited image with the target text and original background.

\subsubsection{Background Inpainting Subnet}
\label{subsubsec:Background}

Prior to performing text editing, we need to erase the text in the original text region and fill the background.
In this part, we adopt the original encoder-decoder structure in SRNet \cite{wu2019editing} to complete the background inpainting.
The $\mathcal{L}_1$ loss and adversarial loss \cite{zhou2018non} is employed to optimize the initial background inpainting subnet.
The loss function of background inpainting subnet written as
\begin{align}
\label{eqn:LossBackground}
\begin{split}
\mathcal{L}_b = &\mathbb{E} \big[ \log D_b (I_b, I_s) + \log (1 - D_b (O_b, I_s)) \big] \\  &+ \lambda _ b \| I_b - O_b \|_1,
\end{split}
\end{align}
where $\mathbb{E}$ denotes the expectation operation, $D_b$ denotes the discriminator network of the background inpainting subnet, $O_b$ is the output of the background inpainting subnet, $I_b$ is the ground-truth of background images, $\lambda_b$ is the weighting factor that is set to 10 to balance adversarial loss and $\mathcal{L}_1$ loss in our experiment.

As shown in Fig.~\ref{fig:BG_Limitation}, the background inpainting performance degrades seriously under complex backgrounds.
As discussed in Sec.~\ref{sec:Literature Review}, the texture continuity in the background region was not considered in the existing network design \cite{wu2019editing, yang2020swaptext}.
In our approach, we adopt the background inpainting subnet in SRNet for a rough reconstruction, and the fine details of background inpainting will be reconstructed in the fusion subnet (Sec.~\ref{subsubsec:Fusion}).

\subsubsection{Text Conversion Subnet}
\label{subsubsec:Text}

The purpose of the text conversion subnet is to convert the target texts to the style of source texts.
In this subnet, the text properties that can be transferred include fonts, sizes, color, etc.

\begin{table}[t!]
\caption{Summary of Symbol Definitions.}
\label{tab:SymbolDefinition}
\vspace{-0.25cm}
\begin{center}
    \begin{tabular}{ | l | l |}
    \hline
    \multicolumn{2}{|c|}{Symbols in TENet} \\ \hline
    $I_s$  & source image with background and text   \\ \hline
    $I_s'$ & source image without background    \\ \hline
    $I_t$ & target textual image with solid background   \\ \hline
    $O_t$ & output of text conversion subnet in TENet    \\ \hline
    $T_{sk}$ & text skeleton image of the target styled characters   \\ \hline
    $I_{st}$ & the target styled characters with solid background    \\ \hline
    $I_b$   & background image of $I_s$     \\ \hline
    $O_b$ & output of the background inpainting subnet in TENet    \\ \hline
    $O_{cf}$ & output of the coarse fusion subnet     \\ \hline
    $O_{ff}$ & output of the fine fusion subnet   \\ \hline
    $T_e$ & edge map extracted from $I_f$    \\ \hline
    $M_t$ &	binary mask of $I_{st}$    \\ \hline \hline
    \multicolumn{2}{|c|}{Symbols in PCNet} \\ \hline
    $I_o$  & original digital document image \\ \hline
    $I_p$  &  $I_o$ with print-and-capture distortion \\ \hline
    $I_d$  & denoised version of $I_p$ \\ \hline
    $O_{p}$  & output of PCNet  \\ \hline \hline
    \multicolumn{2}{|c|}{Symbols in IHNet} \\ \hline
    $I_h$  & generated halftone image used to training IHNet \\ \hline
    $O_{c}$ & output of CoarseNet in IHNet  \\ \hline
    $I_e$  & edge map extracted from $I_h$ \\ \hline
    $O_{e}$  & output of EdgeNet in IHNet  \\ \hline
    $O_{d}$  & output of DetailNet in IHNet  \\ \hline
    $O_{e}^{d}$  & edge map obtained by feeding $O_{d}$ to EdgeNet  \\ \hline
    \end{tabular}
\end{center}
\end{table}

However, the performance of text conversion subnet in \cite{wu2019editing} degraded significantly (as shown in Fig.~\ref{fig:BG_Limitation}) if the background region of the source image $I_s$ contains complex textures.
Therefore, we propose to isolate the text region from the background texture before carrying out text conversion.
Firstly, the background image $O_b$ is obtained by the background inpainting subnet proposed in Sec.~\ref{subsubsec:Background}.
Secondly, we differentiate the background image $O_b$ and the source image $I_s$ to get the source text image without background $I_s'$.
Due to the subtle differences between $O_b$ and the corresponding ground-truth $I_b$, there will be some residuals in the differential image of $I_s$ and $O_b$.
These residuals can be removed by post-processing operation, such as filtering and binarization, and the source text image without background $I_s'$ is obtained.

The target text image $I_t$ and $I_s'$ are fed into text conversion subnet which follows the encoder-decoder FCN framework.
The network can then convert $I_t$ according to the style of $I_s'$ without interference from the background region.

However, different from the training data provided in \cite{wu2019editing}, our target documents (as shown in Fig.~\ref{fig:ID_examples}) contain a significant amount of Chinese characters which are with more complex structure than that of the English alphabets and Arabic numerals.
Besides, the number of Chinese characters is huge, with more than 3,000 characters in common use.
Therefore, instead of using a ResBlock-based text skeletons extraction subnet in \cite{wu2019editing}, we directly adopt a hard-coded component \cite{saeed2010k3m} for text skeleton extraction in our implementation to avoid unnecessary distortions.
Such designs avoid the training overhead for Chinese characters, though the flexibility of the network is reduced.

Intuitively, the $\mathcal{L}_1$ loss can be applied to train text conversion subnet.
However, without weighting the text and background region, the output of text conversion subnet may leave visible artifacts on character edges.
We proposed to add an binary mask of the target styled text image $M_t$ to weight different components in the loss function.
The loss of the text conversion subnet can be written as
\begin{align}
\label{eqn:LossText}
\begin{split}
\mathcal{L}_t = & |M_t|_0 \cdot M_t \cdot \mathcal{L}_{t_1} + (1 - |M_t|_0) \cdot (1 - M_t) \cdot
\mathcal{L}_{t_1},
\end{split}
\end{align}
where $|M_t|_0$ is the $\mathcal{L}_0$ norm of $M_t$, and $\mathcal{L}_{t_1}$ is the $\mathcal{L}_1$ loss between the output of text conversion subnet $O_t$ and the corresponding ground-truth.
It should be noted that during testing, $T_{sk}$ is replaced with the text skeleton image of the intermediate result $O_t'$ after performing decoding.


\subsubsection{Fusion Subnet}
\label{subsubsec:Fusion}

We use the fusion subnet to fuse the output of the background inpainting subnet $O_b$ and the output of the text conversion subnet $O_t$.
In order to improve the quality of the text editing image, we further divide the fusion subnet into coarse fusion subnet and fine fusion subnet.

The coarse fusion subnet follows a generic encode-decode architecture.
We first perform three layers of downsampling of the text-converted output $O_t$.
Next, the downsampled feature maps are fed into 4 residual blocks (ResBlocks) \cite{he2016deep}.
It is noteworthy that we connect the feature maps of the background inpainting subnet to the corresponding feature map with the same resolutions in the decoding layers of coarse fusion subnet to allow a straight path for feature reusing.
After decoding and up-sampling, the coarse fusion image $O_{cf}$ is obtained.
The loss function of the coarse fusion subnet is adopted from SRNet \cite{wu2019editing} as
\begin{align}
\label{eqn:LossCoarseFusion}
\begin{split}
\mathcal{L}_{cf} = &\mathbb{E} \big[ \log D_f (I_f, I_t) + \log (1 - D_f (O_{cf}, I_t)) \big] \\  &+ \lambda _ {cf} \| I_f - O_{cf} \|_1,
\end{split}
\end{align}
where $D_f$ denotes the discriminator network of the coarse fusion subnet, $I_f$ is the ground-truth, $O_{cf}$ is the output of the coarse fusion subnet, and $ \lambda_{cf} $ is the balance factor which is set to 10 in our implementation.

Next, we further improve the quality by considering the continuity of background texture in the fine fusion subnet.
The input to this subnet is a single feature tensor which is obtained by concatenating the coarsely fused image $O_{cf}$ and the edge map $T_e$ along the channel-axis, that is $[O_{cf},  T_e]^T$.
It should be noted that $T_e$ is extracted from the ground-truth using Canny edge detector in the training process; while, in the testing process, $T_e$ is the edge map extracted from output of the coarse fusion subnet $O_{cf}$.


In fine fusion subnet, the edge map of ground-truth plays a role in correcting the detail in the background area and maintaining texture continuity \cite{kim2018deep}.
We attaches $[O_{cf},  T_e]^T$ to 4 ResBlocks to enhance the high-frequency details in the image and to remove the artifacts created by the low-frequency reconstruction in coarse fusion subnet.
The loss function of fine fusion subnet is defined as
\begin{align}
\label{eqn:LossFineFusion}
\mathcal{L}_{ff} = \| I_f - O_{ff} \|_1,
\end{align}
where $O_{ff}$ is the output of the fine fusion subnet.

In order to reduce perceptual image distortion, we introduce a VGG-loss based on VGG-19 \cite{simonyan2014very}.
The VGG-loss is divided into a perceptual loss \cite{johnson2016perceptual} and a style loss \cite{gatys2016image} , which are
\begin{align}
\label{eqn:LossFusionVGG}
\mathcal{L}_{vgg} &= \lambda_{g_1} \cdot \mathcal{L}_{per} + \lambda_{g_2} \cdot \mathcal{L}_{style}, \\
\mathcal{L}_{per} &= \mathbb{E} \big[ \| \phi_i (I_f) - \phi_i (O_{cf}) \| _ 1 \big], \\
\mathcal{L}_{style} &= \mathbb{E} \big[ \| G_i^{\phi} (I_f) - G_i^{\phi} (O_{cf}) \| _ 1 \big],
\end{align}
where $i \in [1,5]$ indexes the layers from $relu1\_1$ to $relu5\_1$ layer of VGG-19 model, $\phi_i$ is the activation map of the $i$-th layer, $G^{\phi}_i$ is the Gram matrix of the $i$-th layer, and the weighting factors $\lambda _ {g_1}$ and $\lambda _ {g_2}$ are set to 1 and 500, respectively.

The whole loss function for the fusion subnet is defined as
\begin{align}
\label{equ:LossFusion}
\mathcal{L}_f = \mathcal{L}_{cf} + \mathcal{L}_{vgg} + \mathcal{L}_{ff}.
\end{align}

Eventually, the overall loss for TENet can be written as
\begin{align}
\label{equ:LossTENet}
\mathcal{L}_{\mbox{\tiny{TENet}}} = \arg \min_{G}\max_{D_b,D_f}(\mathcal{L}_b+\mathcal{L}_t+\mathcal{L}_f).
\end{align}
where $G$ is the generator of TENet.

\subsection{Pre-Compensation Network (PCNet)}
\label{subsection:PreCompensation}

\begin{figure}[t!]
\centering
\includegraphics[width=0.95\linewidth]{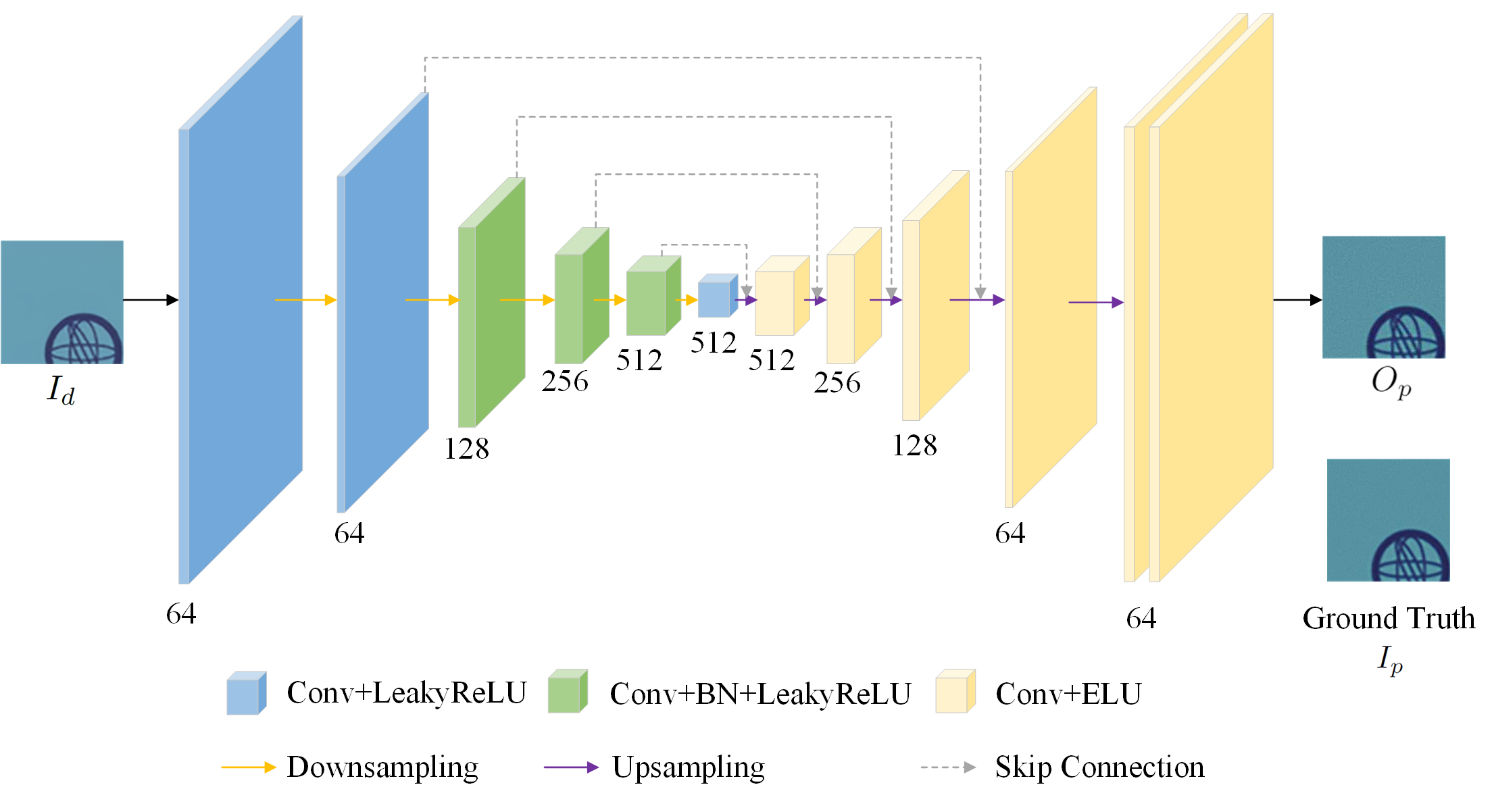}
\caption{Architecture of PCNet. The general architecture follows an encoder-decoder structure.}
\label{fig:ColorCompensation}
\end{figure}

Since the edited text regions are digital images (without print-and-scan distortions), yet the background regions have been through the print-and-scan process.
If stitching the edited text and background regions directly, the boundary artifacts will be obvious.
We propose to pre-compensate the text regions with print-and-scan distortion before combining different regions.
The print-and-scan process introduces nonlinear distortions such as changes in contrast and brightness, various sources of noises, which can be modelled as a non-linear mapping function \cite{zhang2018accurate}.
However, it is more difficult to model the distortion parametrically under uncontrolled conditions.
Inspired by display-camera transfer simulation in \cite{wengrowski2019light}, we propose the PCNet with an auto-encoder structure (shown in Fig.~\ref{fig:ColorCompensation}) to simulate the intensity variation and noise in the print-and-scan process.


We choose the local patch-wise texture matching loss function of the more lightweight VGG-16 network in order to improve the overall performance of the network \cite{kim2018deep}, that is
\begin{align}
\label{eqn:LossTexture}
\mathcal{L}_{tm}(I_p, O_p) = \mathbb{E} \big[ \| G_i^{\phi} (I_p) - G_i^{\phi} (O_p) \| _ 2 \big],
\end{align}
The loss function of PCNet is defined as
\begin{align}
\label{eqn:LossCompensation}
\mathcal{L}_{\tiny\mbox{PCNet}} = \| I_p - O_p \|_1 + \lambda_p \cdot \mathcal{L}_{tm}(I_p, O_p),
\end{align}
where $O_p$ is the output of PCNet, and $I_p$ is the ground-truth of $O_p$.
The local patch-wise texture matching loss between $O_p$ and $I_p$ with weight $\lambda_p$ is also considered.
In our experiment, the weight $\lambda_p$ is set to 0.02.
In practice, the original document image $I_o$ is not accessible by the attacker.
Therefore, a denoised version of the document image $I_d$ is employed in the training process as an estimation of the original document image.
In our experiment, the denoised images are generated by the NoiseWare plugin of Adobe Photoshop \cite{Noiseware}.

Essentially, PCNet learns the intensity mapping and noise distortion in the print-and-scan channel. As shown in Sec.~\ref{subsubsec:AdvancedTamper}, the distortion model can be trained adaptively with a small amount of fine-tuning samples to pre-compensate the channel distortion.


\subsection{Inverse Halftoning Network (IHNet)}

According to \cite{barry1994technique}, halftoning is a technique that simulates the continuous intensity variation in a digital image by changing the size, frequency or shape of the ink dots during printing or scanning.
After the print-and-scan process or processing by our PCNet, the document image can be regarded as clusters of halftone dots.
If the image is re-printed and recaptured without restoration, the halftone patterns generated during the first and second printing process will interfere with each other and introduce aliasing distortions, e.g., Moir\'{e} artifacts \cite{chen2019copy}.
In order to make the forge-and-recapture attack more realistic, the IHNet is proposed to remove the halftoning pattern in the forged document images before recapturing.


We follow the design of network in \cite{kim2018deep} to remove the halftoning dots in the printed document images.
The IHNet can be divided into two steps.
The first step extracts the shape, color (low-frequency features) and edges (high-frequency features) of the document image via CoarseNet and EdgeNet, respectively.
The resulting features are fed into the second stage where image enhancements like recovering missing texture details are implemented.
However, a much simpler structure is adopted since the content of a document image is much more regular and simpler than that of a natural image.
The simplification includes removing the high-level network components (e.g., the object classification subnet) and the discriminator in \cite{kim2018deep}.
By such simplification, the network is much more efficient.

\begin{figure}[t!]
\centerline{\includegraphics[width=0.490\textwidth]{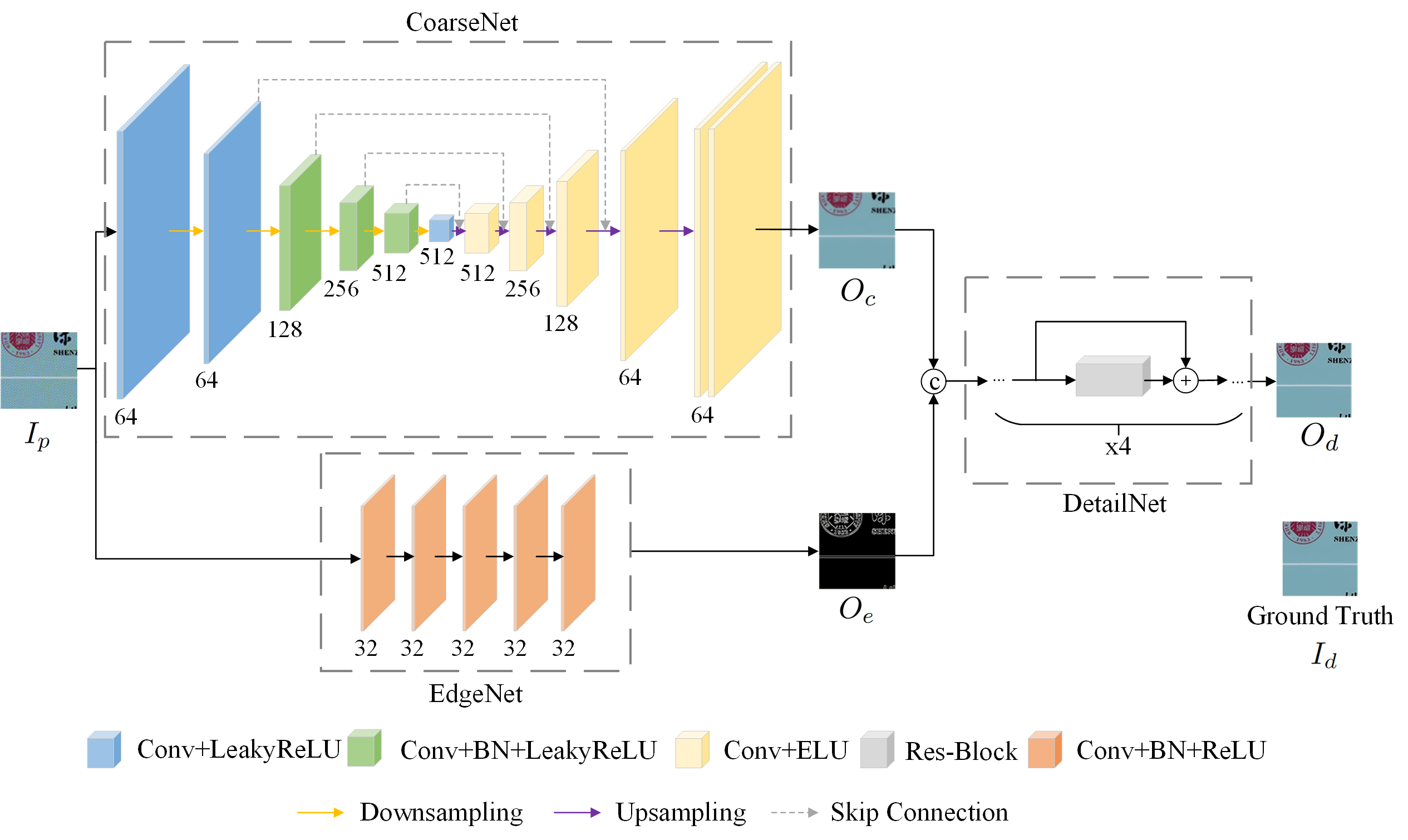}}
\caption{Architecture of IHNet. It consists of three sub-networks: CoarseNet, EdgeNet and DetailNet.}
\label{fig:InverseHalftone}
\end{figure}

Specifically, the CoarseNet with an encoder-decoder structure is employed for the rough reconstruction of the shape and color of halftone input images.
Besides $\mathcal{L}_1$ loss, a global texture loss function (defined in Eq.~\ref{eqn:LossTexture}) based on the VGG-16 structure is used to measure the loss in texture statistics.
Therefore, the overall loss function of CoarseNet is defined as
\begin{align}
\label{eqn:LossInverseHalfton}
\mathcal{L}_{\tiny\mbox{CoarseNet}} = \| I_d- O_c \|_1 + \lambda_c \cdot \mathcal{L}_{tm}(I_d, O_c),
\end{align}
where $O_c$ is the output of CoarseNet and $I_d$ is the denoised version of the document image, and $\lambda_c$ is the weighting factor set to 0.02 in our implementation.

Due to the downsampling operation in the encoder part of CoarseNet, the high-frequency features are not preserved in the reconstructed images.
However, the high frequency components, such as edge and contour of the objects are important visual landmarks in the image reconstruction task.
Therefore, the edge map is provided as auxiliary information to the reconstruction process.

Instead of detecting edges with Canny edge detector (as shown in the fusion subnet in Sec.~\ref{subsubsec:Fusion}), an end-to-end convolutional network is proposed here to extract the contour of characters and background texture from $I_p$.
This is because the traditional edge detector will also detect the edges from halftone dots in $I_p$ which should be removed by the IHNet.
Due to the binary nature of an edge map, the cross-entropy function is used as the loss function of EdgeNet, that is
\begin{align}
\label{eqn:LossEdgenet}
\mathcal{L}_{\tiny\mbox{EdgeNet}} (O_e) = \mathbb{E} \big[- I_e \log (O_e) + ( 1 - I_e ) \log ( 1 - O_e ) \big],
\end{align}
where $I_e$ and $O_e$ are the edge map of the ground-truth and output of EdgeNet, respectively.

The output maps from CoarseNet and EdgeNet are concatenated along the channel-axis to form a single feature tensor before fed into the DetailNet, that is $[O_c, O_e]^T$.
DetailNet adopts the residual network that integrates low and high frequency features.
It clears the remaining artifacts in the low frequency reconstruction, and enhances the details.
The loss function of the network is defined as
\begin{align}
\label{eqn:LossDetail}
\begin{split}
\mathcal{L}_{\tiny\mbox{DetailNet}} = \lambda_{d_1} \| I_d - O_d \|_1 + \lambda_{d_2} \mathcal{L}_{\tiny\mbox{EdgeNet}} (O_e^d) + \lambda_{d_3} \mathcal{L}_{tm} (I_d,O_d),
\end{split}
\end{align}
where $O_d$ is the output of DetailNet and $O_e^d$ is the edge-map obtained by feeding $O_d$ to EdgeNet.
We set the weights as $\lambda_{d_1}=100, \lambda_{d_2}=0.1, \lambda_{d_3}=0.5$, respectively.

\section{Datasets and Training Procedure}
\label{sec:DataAndTraining}

\subsection{Datasets}
\label{subsec:Datasets}

\begin{figure}[t!]
\centering
\subfigure[$I_s$]{
\begin{minipage}[t]{0.2\linewidth}
\centering
\includegraphics[width=0.7in]{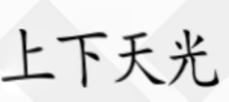}
\end{minipage}
}
\subfigure[$I_t$]{
\begin{minipage}[t]{0.2\linewidth}
\centering
\includegraphics[width=0.7in]{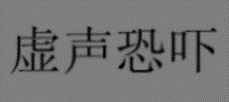}
\end{minipage}
}
\subfigure[$T_{sk}$]{
\begin{minipage}[t]{0.2\linewidth}
\centering
\includegraphics[width=0.7in]{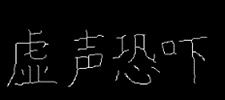}
\end{minipage}
}
\subfigure[$I_{st}$]{
\begin{minipage}[t]{0.2\linewidth}
\centering
\includegraphics[width=0.7in]{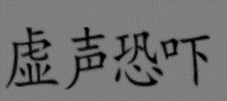}
\end{minipage}
}

\hspace{1mm}\subfigure[$I_b$]{
\begin{minipage}[t]{0.2\linewidth}
\centering
\includegraphics[width=0.7in]{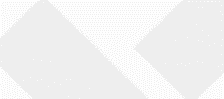}
\end{minipage}
}
\subfigure[$I_f$]{
\begin{minipage}[t]{0.2\linewidth}
\centering
\includegraphics[width=0.7in]{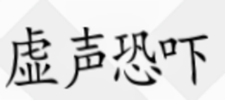}
\end{minipage}
}
\subfigure[$T_e$]{
\begin{minipage}[t]{0.2\linewidth}
\centering
\includegraphics[width=0.7in]{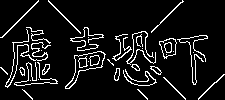}
\end{minipage}
}
\subfigure[$M_t$]{
\begin{minipage}[t]{0.2\linewidth}
\centering
\includegraphics[width=0.7in]{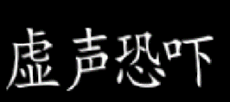}
\end{minipage}
}
\caption{Examples of synthetic Chinese character in dataset $D_t$. (a) $I_s$, source image with styled text and background. (b) $I_t$, target characters. (c) $T_{sk}$, text skeleton of $I_t$. (d) $I_{st}$, target styled characters w/o background. (e) $I_b$, background image. (f) $I_f$, target styled characters with background. (g) $T_e$, edge map from $I_f$. (h) $M_t$, binary mask of $I_{st}$.}
\label{fig:DatabaseDt}
\end{figure}

%
\subsubsection{Synthetic Character Dataset}
\label{subsubsec:ChineseDataset}

The editing object of our task contains a large number of Chinese characters.
To train TENet, we construct a synthetic character dataset $D_t$ including text types in Chinese characters, English alphabets and Arabic numerals.
As shown in Fig.~\ref{fig:DatabaseDt}, the dataset consists of eight types of images, which are summarized as follows:

\begin{itemize}
\item $I_s$: a source image which consists of a background image and generated characters with random content and length, including Chinese characters (about 5 characters per image), English alphabets (about 10 alphabets per image) and Arabic numerals (about 10 alphabets per image), and the colors, fonts and rotation angles are also randomly determined.
\item $I_t$: a gray background image with fixed font for the target character(s).
\item $T_{sk}$: a text skeleton image of $I_t$.
\item $I_{st}$: target styled character(s) with gray solid background.
\item $I_b$: the background image in the source image.
\item $I_f$: an image consisting of both the background of the source image and the target styled character(s).
\item $T_e$: the edge map extracted from $I_f$.
\item $M_t$: the binary mask of $I_{st}$.
\end{itemize}

The synthetic text dataset $D_t$ contains a total of 400,000 images, with 50,000 images of each type.


\subsubsection{Student Card Image Dataset}
\label{subsubsec:HQDataset}

\begin{figure}[t!]
\centerline{\includegraphics[width=3.5in]{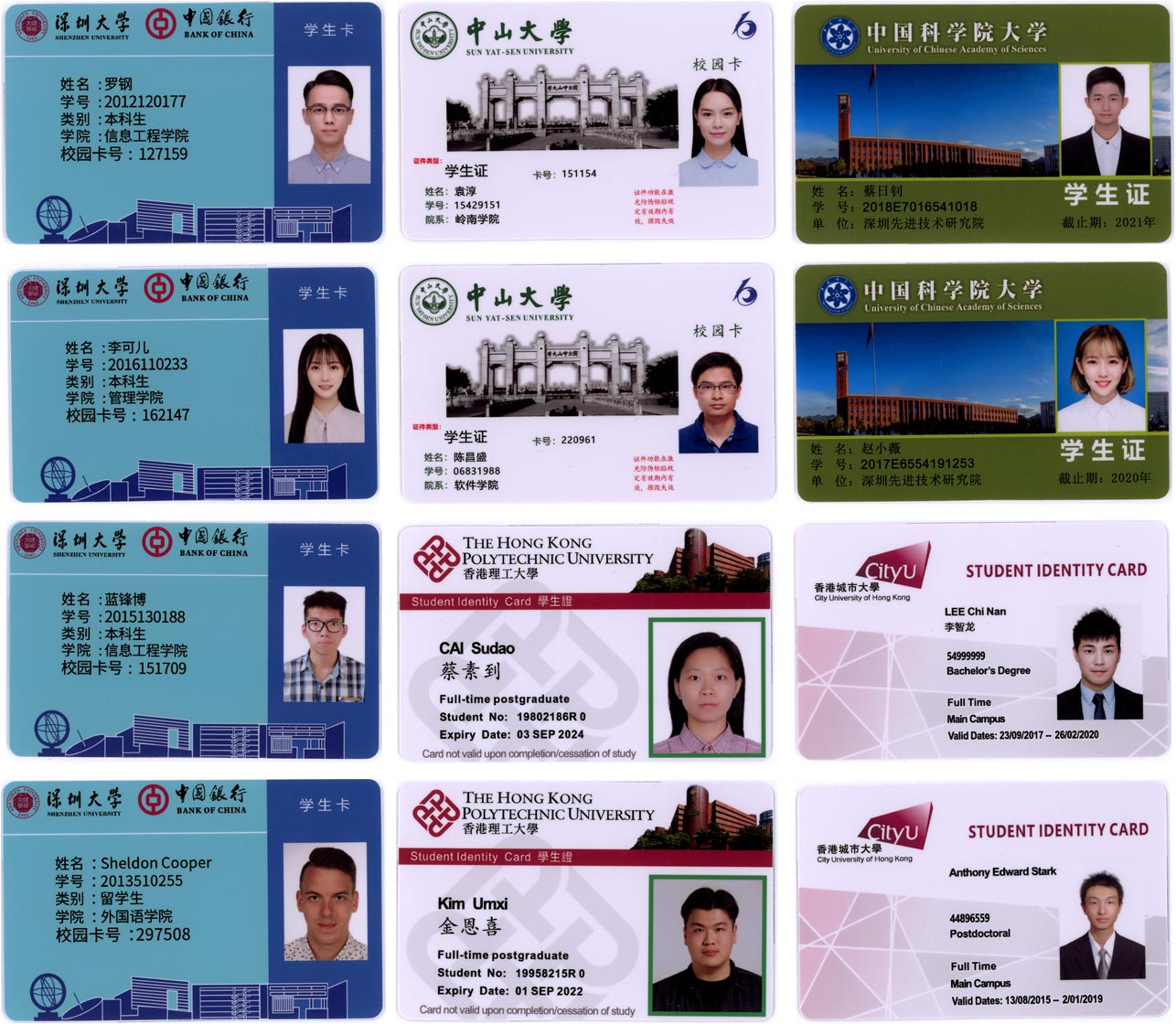}}
\caption{Some captured images of student cards for our experiment. The originals are synthesized with Adobe CorelDRAW.}
\label{fig:CapturedID}
\end{figure}

To facilitate the training of our ForgeNet, a high-quality dataset consists of captured document images from various devices is needed.
As shown in Fig.~\ref{fig:CapturedID}, we use the student card dataset from our group \cite{chen2021database}.
The original images in this dataset are synthesized using Adobe CorelDRAW and printed on acrylic plastic material by a third-party manufacturer.
It contains a total of 12 student cards from 5 universities.
The dataset is collected by 11 off-the-shelf imaging devices, including 6 camera phones (XiaoMi 8, RedMi Note 5, Oppo Reno, Huawei P9, Apple iPhone 6 and iPhone 6s) and 5 scanners (Brother DCP-1519, Benq K810, Epson V330, Epson V850 and HP Laserjet M176n).
In total, the dataset consists of 132 high-quality captured images of student card images.
In our experiments, these document images are used in the forgery and recapture operations.
This dataset is denoted as $D_c$.

\subsection{Training Procedure of ForgeNet}
\label{subsec:TrainingProcedure}

The training process of the proposed ForgeNet is carried out in several phases. The TENet, PCNet and IHNet are pre-trained separately.

\subsubsection{Training TENet}
\label{susbsubsec:TENet}

For training TENet, we use the synthetic chinese character dataset $D_t$ in Sec.~\ref{subsubsec:ChineseDataset}.
In order to cater for the network input dimension, we adjust the height of all images to 128 pixels and keep the original aspect ratio.
In addition, the 400,000 images in the dataset are divided into training set, validation set and testing set in an 8:1:1 ratio.
Different portions of the dataset are fed into the corresponding inputs of the network for training.
With a given training dataset, the model parameters (random initialization) are optimized by minimizing the loss function.
We implement a pix2pix-based network architecture \cite{isola2017image} and train the model using the Adam optimizer ($\beta_1=0.5, \beta_2=0.999$).
The batch size is set to 8.
Since it is not simple to conduct end-to-end joint training on such a complicated network, we first input the corresponding images into the background inpainting subnet and text conversion subnet for pre-training with a training period of 10 epochs.
Subsequently, the fusion subnet joins the end-to-end training with a training period of 20 epochs, and the learning rate gradually decreases from $2 \times 10^{-4}$ to $2 \times 10^{-6}$.
We use a NVIDIA TITAN RTX GPU card for training with a total training duration of 3 days.

\subsubsection{Training PCNet}
\label{subsubsec:PCNet}

\begin{figure}[t]
\centering
\subfigure[]{
\centering
\includegraphics[width=0.8in]{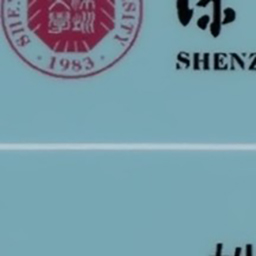}
}
\quad
\subfigure[]{
\centering
\includegraphics[width=0.8in]{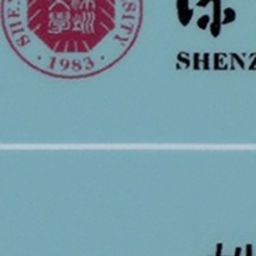}
}
\caption{Illustration of training dataset in PCNet. (a) $I_d$, the denoised version of the captured document image. (b) $I_p$, the captured document image.}
\label{fig:SamplePCNet}
\end{figure}

As shown in Fig.~\ref{fig:SamplePCNet}, training PCNet requires pairs of the original and the captured document images.
PCNet learns the mapping from the original image to the captured image to simulate the print-and-scan distortions.
We use dataset $D_c$ in Sec.~\ref{subsubsec:HQDataset} to train PCNet.
The dataset $D_c$ consists of 12 original images $I_o$ and 132 captured images $I_p$ of the documents.
One may employ $I_o$ and $I_p$ to train PCNet so as to learn the distortions in print-and-scan channel.
However, in practice, it is often difficult for an attacker to obtain the original document image $I_o$.
Alternatively, we use the denoised version of the captured document images $I_d$ as an approximation of the original images.
In our experiment, the NoiseWare plugin in Adobe Photoshop \cite{Noiseware} is employed to remove the distortions in the captured images.
In order to accommodate the network input size, all images in the dataset $D_c$ are cropped to image patches with a resolution of $256 \times 256$ pixels.
In addition, data augmentation strategies such as rotation, cropping, and mirroring are carried out to expand the number of datasets $D_c$ to 20,000 image patches, with 80\% of the data used for training, 10\% for validation, and 10\% for testing.
PCNet is trained with 20 epochs using the ADAM optimizer ($\beta_1=0.9, \beta_2=0.999$) with a learning rate of $1\times10^{-4}$ and no weight decay and the batchsize is set to 8.
The parameter of activation functions LeakyReLu and ELU are set as $\alpha = 0.2$ and $\alpha = 1.0$, respectively.
The training process lasts for 1 day on a NVIDIA TITAN RTX GPU card.

\begin{figure}[t]
\centering
\subfigure[]{
\begin{minipage}[t]{0.25\linewidth}
\centering
\includegraphics[width=0.8in]{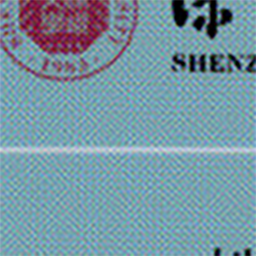}
\end{minipage}
}
\subfigure[]{
\begin{minipage}[t]{0.25\linewidth}
\centering
\includegraphics[width=0.8in]{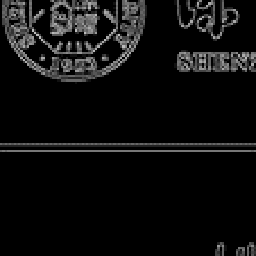}
\end{minipage}
}
\subfigure[]{
\begin{minipage}[t]{0.25\linewidth}
\centering
\includegraphics[width=0.8in]{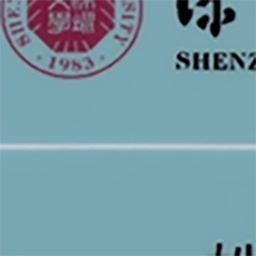}
\end{minipage}
}
\caption{Illustration of training dataset in IHNet. (a) $I_h$, the generated halftone image. (b) $I_e$, the edge image. (c) $I_d$, the denoised version of the captured document image.}
\label{fig:SampleIHNet}
\end{figure}

\subsubsection{Training IHNet}
\label{subsubsec:IHNet}

\begin{figure*}[t!]
\centering
\subfigure[Original]{
\includegraphics[width=0.12\textwidth]{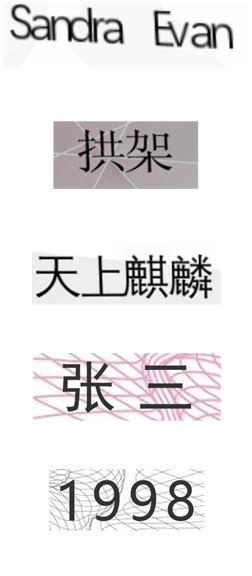}
}
\subfigure[SRNet \cite{wu2019editing}]{
\includegraphics[width=0.12\textwidth]{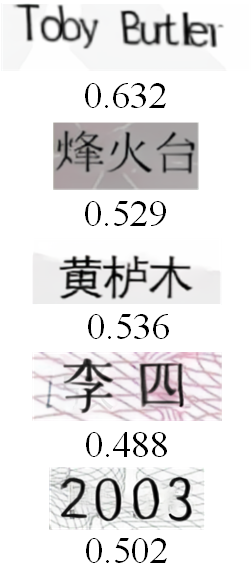}
\label{subfig:SRNet}
}
\subfigure[TENet w/o ID]{
\includegraphics[width=0.12\textwidth]{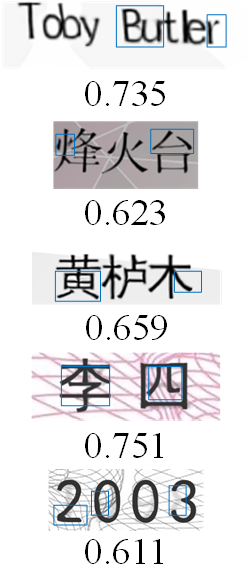}
\label{subfig:woid}
}
\subfigure[TENet w/o FF]{
\includegraphics[width=0.12\textwidth]{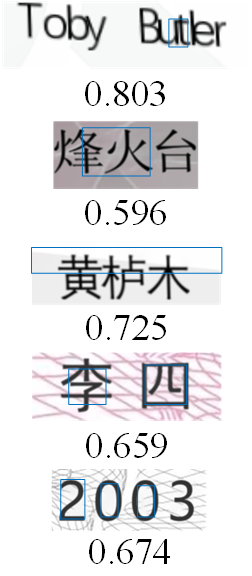}
\label{subfig:woff}
}
\subfigure[TENet w/o SS]{
\includegraphics[width=0.12\textwidth]{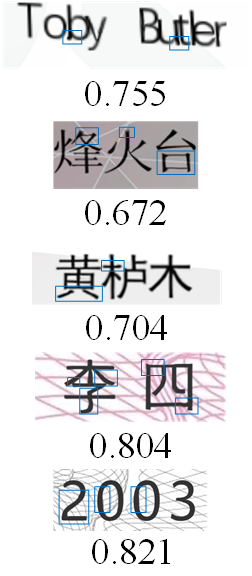}
\label{subfig:wosk}
}
\subfigure[TENet]{
\includegraphics[width=0.12\textwidth]{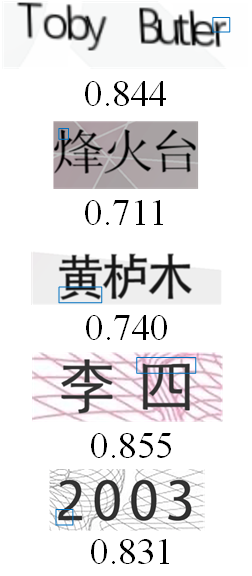}
\label{subfig:TENetfull}
}
\subfigure[Ground-truth]{
\includegraphics[width=0.12\textwidth]{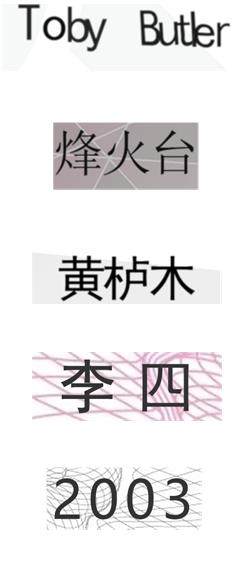}
}
\caption{Comparisons of SRNet \cite{wu2019editing} and different configurations of the proposed TENet on synthetic character dataset $D_t$. (a) Original images. (b) Edited by SRNet \cite{wu2019editing}. (c) Edited by TENet without image differentiation (ID). (d) Edited by TENet without fine fusion (FF). (e) Edited by TENet without skeleton supervision (SS). (f) Edited by the proposed TENet. (g) Ground-truth. Differences between the results from TENet and the ground-truth are boxed out in blue. The SSIM metric computed from each edited document and the ground-truth is shown under each image from (b) to (f).}
\label{fig:AS_TENet}
\end{figure*}

As shown in Fig.~\ref{fig:SampleIHNet}, the denoised document image $I_d$ in dataset $D_c$ is also used to train IHNet.
The edge image $I_e$ of $I_d$, and the artificially generated halftone image $I_h$ are also needed in the training process.
The halftone image $I_h$ is generated by applying color halftone patterns to $I_d$ in Photoshop with amplitude modulation technique and various parameters (random halftone angles for different color channels).
The edge image $I_e$ is an edge map of $I_d$ obtained by Canny edge detection.
Similarly, all images in the dataset are cropped to a resolution of $256 \times 256$ pixels to fit the size of the network input.
Data augmentation strategies are also employed to expand the dataset to 20,000 images, which are then divided into ratio of 8:1:1 for the training, validation and testing sets, respectively.
IHNet uses an ADAM optimizer ($\beta_1=0.9, \beta_2=0.999$) with an initial learning rate of $1\times10^{-4}$.
Since the network is divided into three subnets, we first pre-train CoarseNet and EdgeNet, respectively.
After 10 epochs, DetailNet joins the end-to-end training with a decaying rate of 0.9.
The batchsize is set to 8.
The training stops after 20 epochs.
The training lasts for 2 days on a NVIDIA TITAN RTX GPU card.

\section{Experimental Results}
\label{sec:Experiment}

In the following, we first evaluate the performance of the proposed TENet in both the synthetic character dataset and the student card dataset without distortions from the print-and-scan channel. Then, the performance of ForgeNet (including TENet, PCNet and IHNet) is studied under practical setups, including forgery under the channel distortion, with a single sample, and attacking the state-of-the-art document authentication systems. Finally, some future research directions on detection of such forge-and-recapture attack are discussed.

\subsection{Performance Evaluation on TENet}
\label{subsec:Comparison}

\subsubsection{Performance on Synthetic Character Dataset}
\label{subsubsec:Ablation}

In Sec.~\ref{subsec:TextEdit}, we propose the text editing network, TENet by adapting SRNet \cite{wu2019editing} to our task.
However, SRNet is originally designed for editing English alphabets and Arabic numerals in scene images for visual translation and augmented reality applications.
As shown in Fig.~\ref{fig:BG_Limitation}, \ref{fig:Chinese_Limitation} and \ref{subfig:SRNet}, it does not perform well on Chinese characters with complicated structure, especially in document with complex background.
In this part, we qualitatively and quantitatively examine the modules in TENet which are different from SRNet so as to show the effectiveness of our approach.
Three main differences between our proposed SRNet and TENet are as follows.
First, we perform image differentiation operation between the source image $I_s$ and the output $O_b$ of the background inpainting subnet to obtain style text image without background $I_s'$.
Second, $I_s'$ is then fed into a hard-coded component to extract the text skeleton of the style text which is then directly input to the text conversion subnet as supervision information.
Third, instead of only using a general U-Net structure to fuse different components (as in SRNet), we adopt a fine fusion subnet in TENet with consideration on texture continuity.
We randomly select 500 images from our synthetic character dataset $D_t$ as a testing set for comparison.
Quantitative analysis with three commonly used metrics are performed to evaluate the resulting image distortion, including Mean Square Error (MSE, a.k.a. $l_2$ error), Peak Signal-to-Noise Ratio (PSNR), and Structural Similarity (SSIM) \cite{wang2004image}.
The edited results by different approaches are compared with the ground-truth to calculate these metrics.

\begin{table}[t!]
\centering
\caption{Comparisons of SRNet \cite{wu2019editing} and different settings of TENet. The best results are highlighted in bold.}
\label{tab:TENet}
\begin{tabular}{cccc}
\hline
Method & MSE & PSNR & SSIM \\
\hline
SRNet\cite{wu2019editing} & 0.032 & 16.44 & 0.519 \\
\hline
TENet w/o ID & 0.027 & 17.37 & 0.687\\
TENet w/o FF & 0.019 & 19.14 & 0.635\\
TENet w/o SS & 0.015 & 19.75 & 0.708\\
TENet & \textbf{0.011} & \textbf{20.48} & \textbf{0.731}\\
\hline
\end{tabular}
\end{table}

\textbf{Image Differentiation (ID).}
After removing the image differentiation part, we find that the text generation gets worse (as shown in Fig.~\ref{subfig:woid}).
The distortion is more severe in the case of source images with complex backgrounds.
Due to the interference of the background textures, text conversion subnet cannot accurately distinguish the foreground (characters) from background.
It leads to difficulty in extracting text style features and character strokes are therefore distorted.
For example, the residual of the original characters are still visible in background of the last two figures in Fig.~\ref{subfig:woid}).
In contrast, using the differential image of the source image $I_s$ and the output $O_b$ as input to the text conversion subnet can avoid background interference, allowing the text conversion subnet to focus on extracting text styles without confusions from the background.
It leads to a better text conversion performance.
From Tab.~\ref{tab:TENet}, we can see that without image differentiation, there is a significant drop in MSE and PSNR compared to the proposed TENet.
The above experiments indicate that the differentiation operation is essential in producing high quality styled text.

\textbf{Fine Fusion (FF).}
The performance of TENet under complex background mainly relies on the fine fusion subnet.
If the fine fusion subnet is removed, the resulting image suffers from loss of high-frequency details.
This is because the remaining subnets (the background inpainting subnet and the text conversion subnet) are of U-Net based structure which downsamples the input images before reconstruction.
As shown in Fig.~\ref{subfig:woff}, the resulting text images are blurry.
Besides, the SRNet does not take into account the continuity of the background texture during image reconstruction.
The texture components in the resulting images are discontinuous.
The results of Tab.~\ref{tab:TENet} show that the impact of removing fine fusion component is much more significant than the others.
It is due to the fact that the background region is much larger than the foreground region in our test images, and the contribution of the fine fusion subnet is mainly in the background.

\textbf{Skeleton Supervision (SS).}
Visually, Chinese characters are much more complex than English alphabets and Arabic numerals in terms of the number of stokes and the interaction of the strokes in a character.
The skeleton supervision information is important in providing accurate supervision on the skeleton of Chinese characters.
If the skeleton is extracted using a general trainable network (as designed in SRNet) instead of using a hard-coded the style text, the text skeleton extraction performance will be degraded.
As shown in Fig.~\ref{subfig:wosk}, by removing the skeleton supervision component, the character strokes in the resulting images appear distorted and the characters are not styled correctly.
From Tab.~\ref{tab:TENet}, we learn that the skeleton supervision has less impact on the overall image quality, as it only affects the character stroke generation.
However, the style of characters is vital in creating high quality forgery samples.

In summary, the results look unrealistic in the absence of these three components as shown in the ablation study in Fig.~\ref{fig:AS_TENet}(c)-(e).
The importance of image differentiation, fine fusion, and skeleton supervision are reflected in the quality of characters, the background texture, and the character skeleton, respectively.
Both quantitative analysis and visual examples clearly indicate the importance of the three components.

\begin{figure}[t!]
\centering
\subfigure[Original]{
\centering
\includegraphics[width=0.11\textwidth]{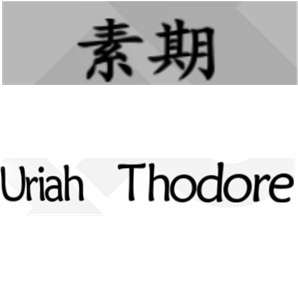}
}\hspace{-3mm}
\subfigure[Target text]{
\centering
\includegraphics[width=0.11\textwidth]{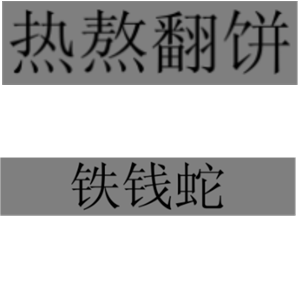}
}\hspace{-3mm}
\subfigure[TENet]{
\centering
\includegraphics[width=0.11\textwidth]{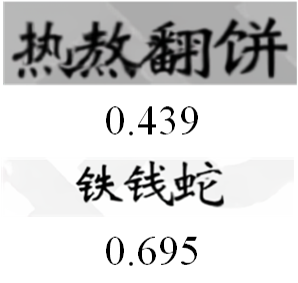}
}\hspace{-3mm}
\subfigure[Ground-truth]{
\centering
\includegraphics[width=0.11\textwidth]{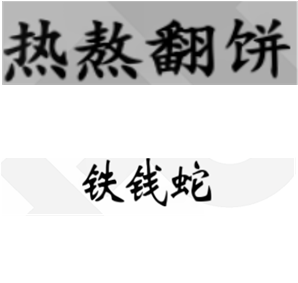}
}
\caption{Some failure cases. The SSIM metric computed from edited document and the ground-truth is shown under the image (c).}
\label{fig:FailExample}
\end{figure}

Although TENet shows excellent text editing performance on most document images, it still has some limitations. When the structure of target character is complex or the number of characters is large, TENet may fail. Fig.~\ref{fig:FailExample} shows two failure cases. In the top row, the performance of the text conversion subnet is degraded due to the complex structure and large number of strokes of the target characters, and thus the editing results show distortion of the strokes. In the bottom row, it is a text conversion with cross languages and different character lengths. In dataset $D_t$, we follow the dataset generation strategy of SRNet \cite{wu2019editing}, where source and target styled characters have the same geometric attribute (e.g., size, position) settings. However, for pairs of characters of different lengths, the strategy for setting the text geometry attributes is to make the overall style of the text with fewer characters converge to that of multiple characters. But inevitably, some geometric attributes of text with fewer characters are missing. The text conversion process of TENet excellently implements the conversion of geometric attributes from source text to target styled text, thus causing the generated results to have errors with ground-truth. These failures occur because the number and type of samples in the training data are insufficient, which leads to the unsatisfactory generalization performance of the model. So we believe that these problems could be alleviated by adding more complex characters and more font attributes to the training set.

\begin{figure}[t!]
\centering
\subfigure[Original]{
\centering
\includegraphics[width=0.155\textwidth]{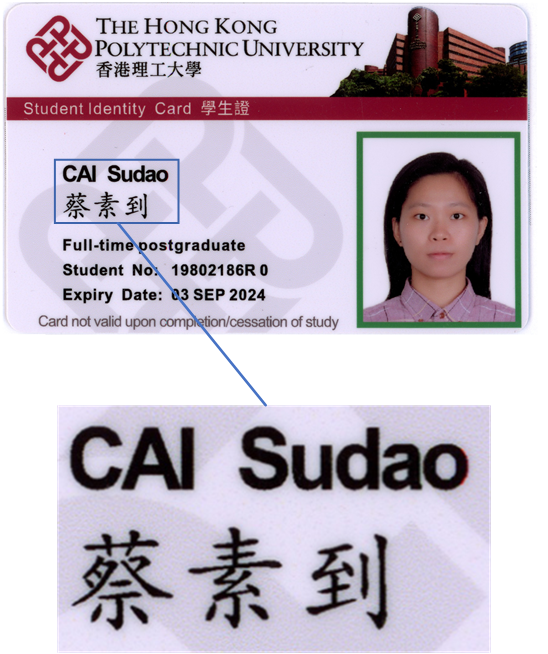}
}\hspace{-3mm}
\subfigure[SRNet \cite{wu2019editing}]{
\centering
\includegraphics[width=0.155\textwidth]{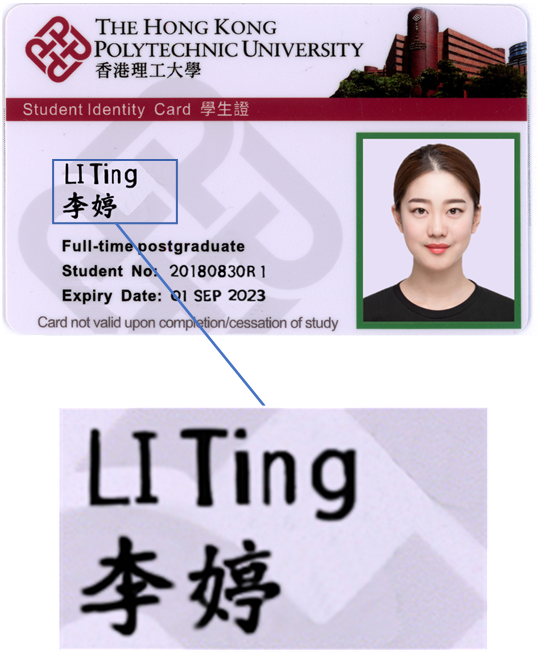}
}\hspace{-3mm}
\subfigure[TENet]{
\centering
\includegraphics[width=0.155\textwidth]{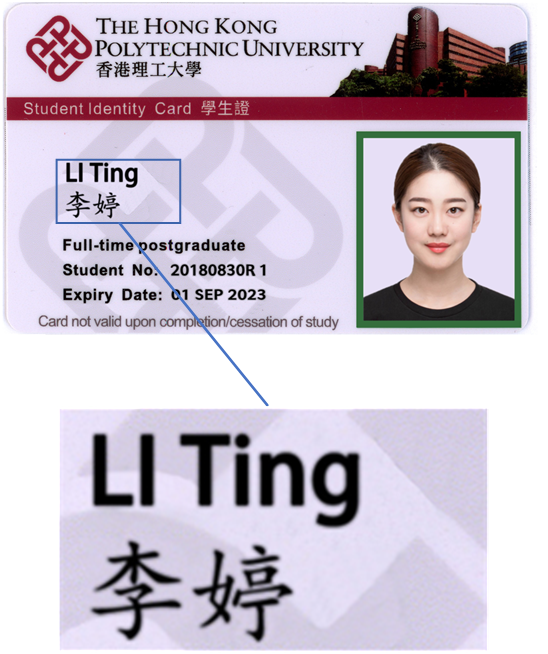}
}
\caption{Visual comparison on the images of student cards without print-and-scan distortions. (a) Original image. (b) Edited by SRNet \cite{wu2019editing}. (c) Edited by the proposed TENet.}
\label{fig:VC}
\end{figure}

\subsubsection{Performance on the Student Card Forgeries}
\label{subsubsec:VisualComparison}

In Sec.~\ref{subsubsec:Ablation}, we perform an ablation study of the text editing module in a target text region of the document.
However, it has not reflected the forgery performance on the entire image, including text, image and background as shown in Fig.~\ref{fig:ForgeNet}.
In this part, we perform text editing on the captured student card images and stitch the edited text regions with the other regions to yield the forged document image.
It should be noted that the print-and-scan distortion is not considered in this experiment since we are evaluating the performance of TENet.

In this experiment, SRNet \cite{wu2019editing} and the proposed TENet are compared in the text editing task with some student cards of different templates from dataset $D_c$.
The training data contains 50,000 images from each type of images introduced in Sec.~\ref{subsec:Datasets}-1).
The height of all images is fixed to 128 pixels, and the original aspect ratio is maintained.
The edited text fields are name, student No. and expiry date, including Chinese characters, English alphabets and Arabic numerals.
It should be noted that the text lengths may be different before and after editing.
As can be seen from Fig.~\ref{fig:VC}, the proposed TENet significantly improves the performance in character style conversion.

\subsection{Performance Evaluation on ForgeNet}
\label{subsec:Performance}

\subsubsection{Ablation Study of PCNet and IHNet}
\label{subsubsec:TamperResult}

\begin{figure}[t!]
\centering
\subfigure[ForgeNet]{
\centering
\includegraphics[width=0.22\textwidth]{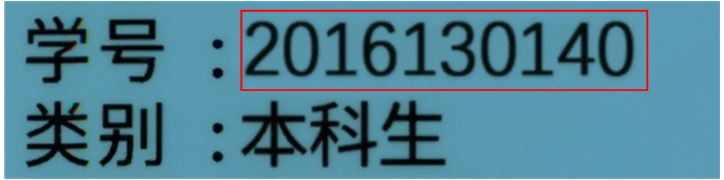}
}
\subfigure[ForgeNet w/o PCNet \& IHNet]{
\centering
\includegraphics[width=0.22\textwidth]{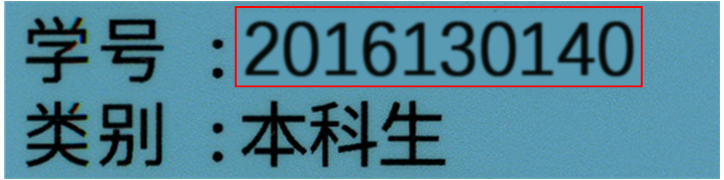}
}

\hspace{0.25mm}
\subfigure[ForgeNet w/o IHNet]{
\centering
\includegraphics[width=0.22\textwidth]{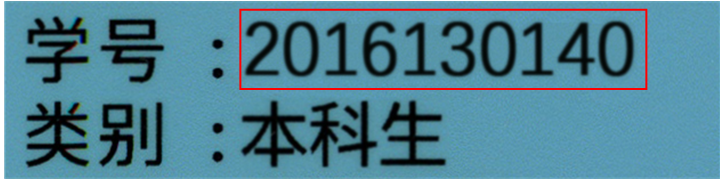}
}
\subfigure[ForgeNet w/o PCNet]{
\centering
\includegraphics[width=0.22\textwidth]{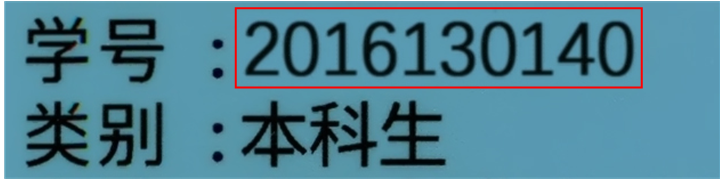}
}
\caption{Examples of ablation study on ForgeNet. (a) The proposed ForgeNet (TENet+PCNet+IHNet). (b) TENet only (w/o PCNet \& IHNet). (c) w/o IHNet. (d) w/o PCNet. The editing regions are boxed out in red.}
\label{fig:AS_PCNet_and_IHNet}
\end{figure}

This part shows the tampering results of ForgeNet under print-and-scan distortion.
The ForgeNet consists of three modules, namely, TENet, PCNet, and IHNet.
We perform ablation study to analyze the role of each module.

The role of the TENet is to alter the content of text region.
However, as shown in Fig.~\ref{fig:AS_PCNet_and_IHNet}(b), the resulting text regions from TENet are not consistent with the surrounding pixels.
This is because the edited region has not been through the print-and-scan channel.
The main channel distortion includes color difference introduced by illumination conditions, different color gamuts and calibration in different devices, as well as halftoning patterns.

One of the most significant difference is in color because printing and scanning process are with different color gamut, and the resulting color will thus be distorted.
Another difference is on the micro-scale in the image which is introduced by the halftoning process and various source of noise in the print-and-scan process.
Thus, the role of PCNet is to pre-compensate the output images with print-and-scan distortions.
As shown in Fig.~\ref{fig:AS_PCNet_and_IHNet}(c), both the edited and background regions are more consistent after incorporating the PCNet.
However, the halftoning artifacts (visible yellow dots) remains.
The remaining halftoning artifacts interfere with the halftoning patterns which happens in the recapturing (print-and-capture) process.
Thus, IHNet removes the visible halftoning artifacts (as shown in Fig.~\ref{fig:AS_PCNet_and_IHNet}(a) and (d)) before performing recapturing attack.
The resulting image processed with both PCNet and IHNet is closer to the original image, which shows that all three modules in ForgeNet play important roles.


\begin{figure}[t!]
\centering
\subfigure[Original]{
\centering
\includegraphics[width=0.15\textwidth]{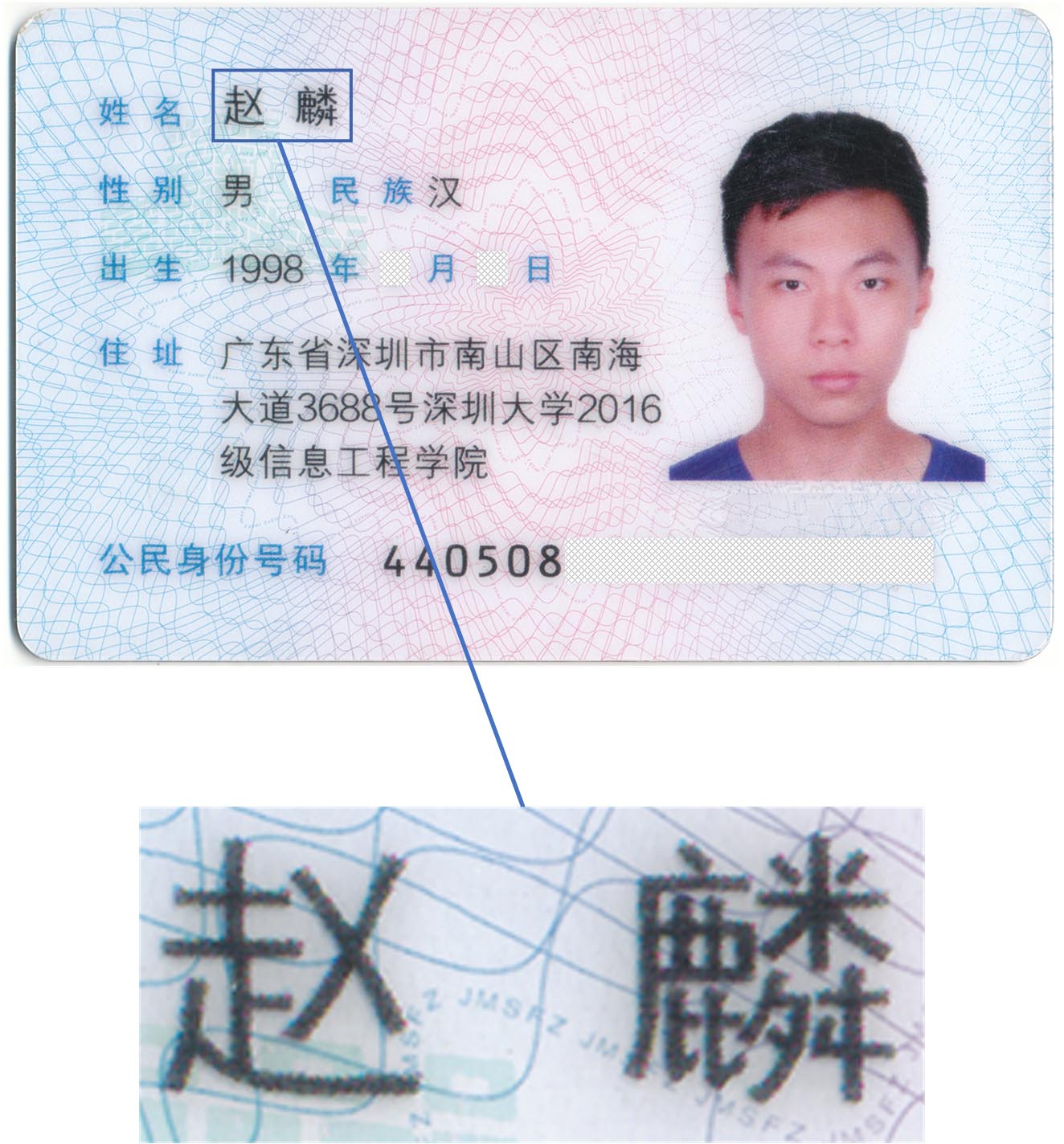}
}\hspace{-3mm}
\subfigure[SRNet \cite{wu2019editing}]{
\centering
\includegraphics[width=0.15\textwidth]{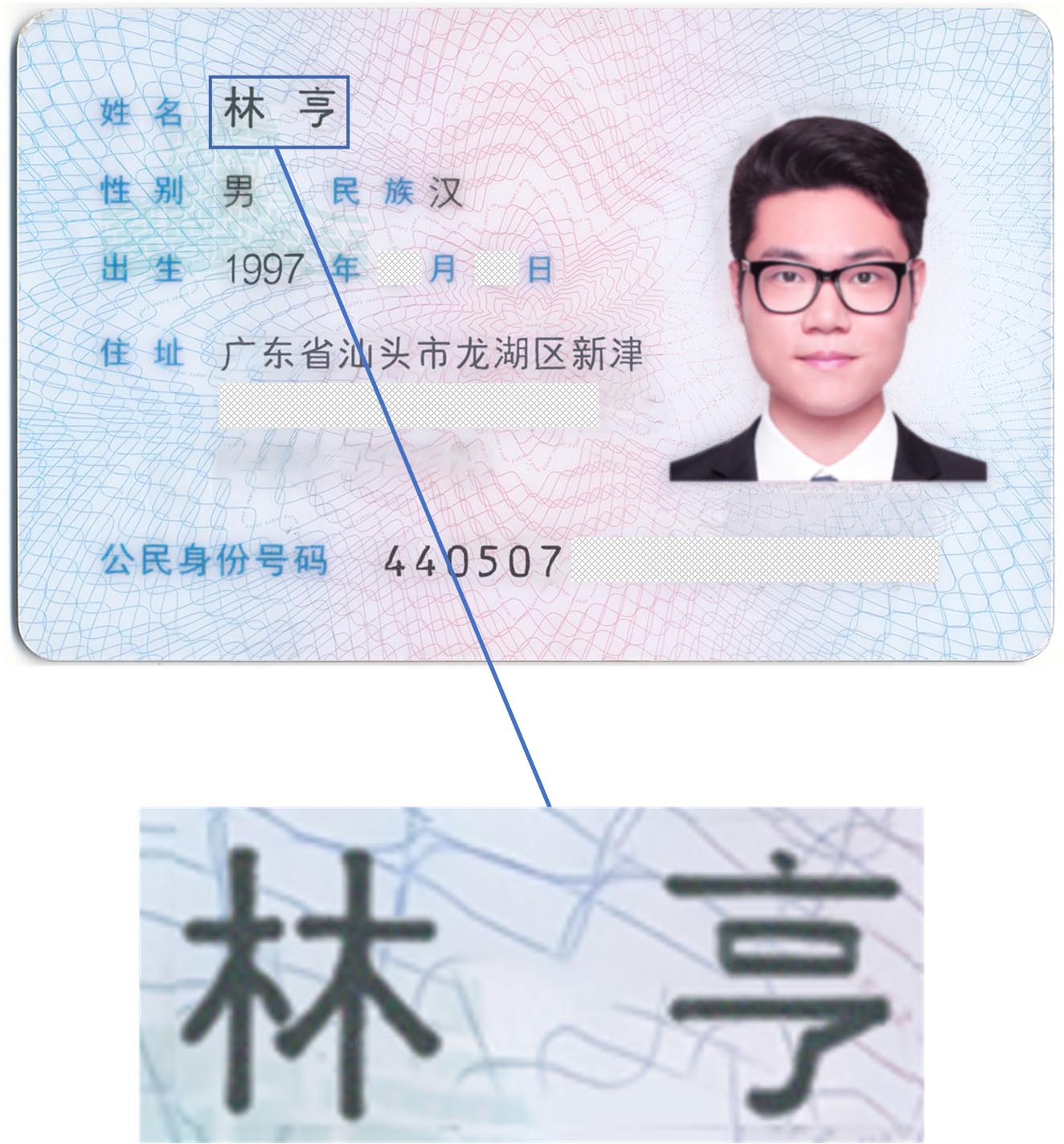}
}\hspace{-3mm}
\subfigure[TENet]{
\centering
\includegraphics[width=0.15\textwidth]{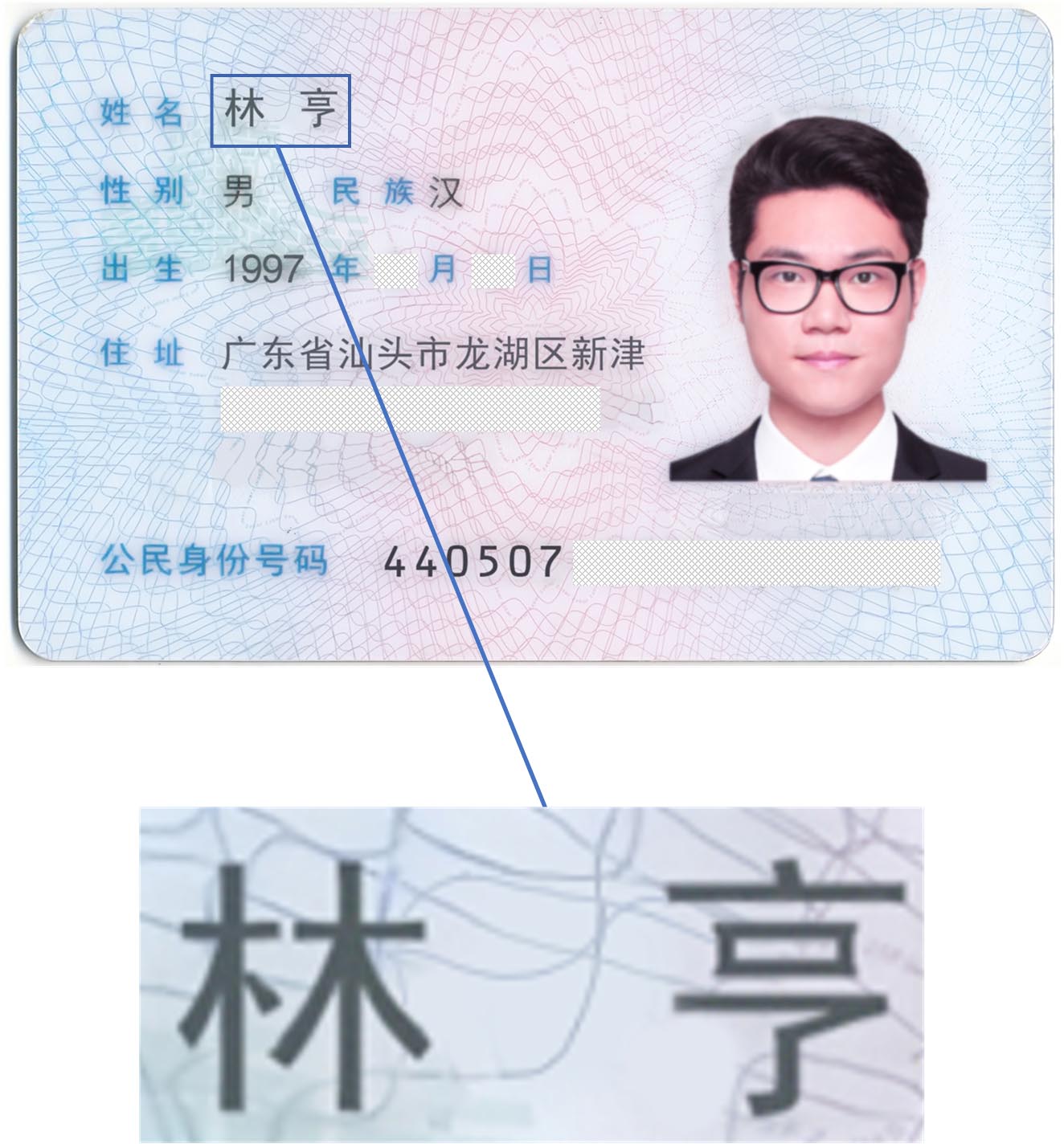}
}
\caption{Visual comparison on the identity card images. (a) Original image. (b) Edited by SRNet \cite{wu2019editing}. (c) Edited by the proposed TENet. Both SRNet and TENet are fine-tuned with a single sample.}
\label{fig:SingleSample}
\end{figure}

\subsubsection{Document Forgery with a Single Sample}
\label{subsubsec:AdvancedTamper}

In the previous section, we show the performance of the proposed ForgeNet on editing student card images.
However, the background regions of these samples are relatively simple, usually with solid colors or simple geometric patterns.
In this part, we choose Resident Identity Card for People's Republic of China with a complex background as a target document.
Identity card tampering is a more practical and challenging task to evaluate the performance of the proposed ForgeNet.
However, identity card contains private personal information.
It is very difficult to obtain a large number of scanned identity cards as training data.
Thus, we assume the attacker has access to only one scanned identity card image which is his/her own copy according to our threat model in Fig.~\ref{fig:TreadModel}(a).
This identity card image is regarded as both the source document image (to be edited) and the sample in target domain for fine-tuning TENet, PCNet and IHNet.
The attacker then tries to forge the identity card image of a target person by editing the text.

The identity card is scanned with a Canoscan 5600F scanner with a resolution of 1200 DPI.
The whole image is cropped according to different network input sizes, and data augmentation is performed.
In total, 5,000 image patches are generated to fine-tune the network.
It is worth noting that the complex textures of the identity card background pose a significant challenge to the text editing task.
To improve the background reconstruction performance, the attacker could include some additional texture images which are similar to the identity card background for fine-tuning.
Some state-of-the-art texture synthesis networks can be employed to generate the texture automatically \cite{xian2018texturegan}.
The image patches are fed to TENet, PCNet, and IHNet for fine-tuning.
In order to collect the sensitive information in identity cards, we need to collect personal information from our research group to finish the forgery test.
Ten sets of personal information (e.g., name, identity number) are gathered for a small-scale ID card tampering test, and 10 forged identity card images are generated accordingly.
As shown in Fig.~\ref{fig:SingleSample}, some key information on the identity card is mosaicked to protect personal privacy.
It is shown that ForgeNet achieves a good forgery performance by fine-tuning with only one image, while the text and background in the image reconstructed by SRNet are distorted.

\begin{table}[t]
\centering
\caption{Identity Document Authentication under Forge-and-Recapture Attack on MEGVIIFace++ AI \cite{Megvii}. The items "Edited", "Photocopy", "Identity Photo" and "Screen" denote the probabilities of image editing, photocopies, identity card images and screen recapturing, respectively.}
\label{tab:FRAA1}
\begin{tabular}{ccccc}
\hline
No. & Edited & Photocopy & Identity Photo & Screen\\
\hline
01 & 0 & 0 & 0.994 & 0.006\\
02 & 0 & 0 & 0.996 & 0.004\\
03 & 0 & 0 & 0.941 & 0.059\\
04 & 0 & 0 & 0.913 & 0.087\\
05 & 0.009 & 0 & 0.991 & 0\\
06 & 0 & 0 & 0.958 & 0.042\\
07 & 0 & 0 & 0.958 & 0.042\\
08 & 0 & 0 & 0.989 & 0.011\\
09 & 0 & 0 & 0.983 & 0.017\\
10 & 0 & 0 & 0.977 & 0.023\\
\hline
\end{tabular}
\end{table}

\subsubsection{Forge-and-Recapture Document Attack Authentication}
\label{subsubsec:Authentication}

In this part, the forged identity card images obtained in Sec.~\ref{subsubsec:AdvancedTamper} are processed by the print-and-scan channel to demonstrate the threat posed by the forge-and-recapture attack.
The printing and scanning devices used for the recapturing process are Canon G3800 and Canoscan 5600F, respectively.
The highest printing quality of $4800 \times 1200$ DPI is employed.
The print substraces is Kodak $230g/m^2$ glossy photo paper.
The scanned images are in TIFF or JPEG formats with scanning resolutions (ranging from 300 DPI to 1200 DPI) adjusted according to the required size of different authentication platforms.

The popular off-the-shelf document authentication platforms in China includes Baidu AI, Tencent AI, Alibaba AI, Netease AI, Jingdong AI, MEGVII Face++ AI, iFLYTEK AI, Huawei AI, etc.
However, the document authentication platforms which detect identity card recapturing and tampering are Baidu AI \cite{Baidu}, Tencent AI \cite{Tencent}, and MEGVII Face++ AI \cite{Megvii}.
We uploaded tampering results to these three state-of-the-art document authentication platforms for validation of the forge-and-recapture identity documents.

The authentication results on MEGVII Face++ AI are shown in Tab.~\ref{tab:FRAA1}.
It is shown that the 10 forge-and-recapture identity images in our test are successfully authenticated.
All the tested images also pass the other two authentication platforms (include inspection against editing, recapturing, etc.).
Given the fact that the state-of-the-art document authentication platforms have difficulties in distinguishing the forge-and-recapture document images, it fully demonstrates the success of our attack.
This calls for immediate research effort in detecting such attacks.

\subsection{Discussion on Detection of Forge-and-Recapture Attack}

As discussed in Section~\ref{sec:Introduction}, the main focus of this work is to build a deep learning-based document forgery network to study the risk of existing digital document authentication system.
Thus, developing forensics algorithm against the forge-and-recapture attack is not in the scope of this work.
Moreover, in order to study such attack, a large and well-recognized dataset of forge-and-recapture document images is needed.
However, no such dataset is currently available in the public domain.
Without such resource, some data-driven benchmarks in digital image forensics with hundreds or thousands feature dimensions \cite{fridrich2012rich, li2016identification} are not applicable.
Meanwhile, this work enables an end-to-end framework for generating high quality forgery document, which facilitates the construction of a large-scale and high-quality dataset.
Last but not least, it has been shown in our parallel work \cite{chen2021database} that the detection of document recapturing attack alone (without forgery) is not a trivial task when the devices in training and testing sets are different.
The performance of generic data-driven approaches (e.g., ResNet \cite{he2016deep}) and traditional machine learning approach with handcrafted features (e.g., LBP+SVM \cite{wen2015face}) are studied.
The detection performance degraded seriously in a cross dataset experimental protocol where different printing and imaging devices are used in collecting the training and testing dataset.

\section{Conclusion}
\label{sec:Conclusion}

In this work, the feasibility of employing deep learning-based technology to edit document image with complicated characters and complex background is studied.
To achieve good editing performance, we address the limitations of existing text editing algorithms towards complicated characters and complex background by avoiding unnecessary confusions in different components of the source images (by the image differentiation component introduced in Sec.~\ref{subsubsec:Text}), constructing texture continuity loss and providing auxiliary skeleton information (by the fine fusion and skeleton supervision components in Sec.~\ref{subsubsec:Fusion}).
Comparisons with the existing text editing approach \cite{wu2019editing} confirms the importance of our contributions.
Moreover, we propose to mitigate the visual artifacts of text editing operation by some post-processing (color pre-compensation and inverse halftoning) considering the print-and-scan channel.
Experimental results show that the consistency among different regions in a document image are maintained by these post-processing.
We also demonstrate the document forgery performance under a practical scenario where an attacker generates an identity document with only one sample in the target domain.
Finally, the recapturing attack is employed to cover the forensic traces of the text editing and post-processing operations.
The forge-and-recapture samples by the proposed attack have successfully fooled some state-of-the-art document authentication systems.
From the study of this work, we conclude that the advancement of deep learning-based text editing techniques has already introduced significant security risk to our document images.

\ifCLASSOPTIONcaptionsoff
  \newpage
\fi

\input{DocumentForgery.bbl}

\bibliographystyle{IEEEtran}

\end{document}